\documentclass[%
  reprint,
  nofootinbib,
  amsmath,amssymb,
  prd,
  floatfix,
  showkeys,
]{revtex4-2}

\usepackage{graphicx}
\usepackage{subfigure}
\usepackage{dcolumn}
\usepackage{bm}
\usepackage{booktabs} 
\usepackage{multirow} 
\usepackage{xcolor} 

\begin{document}

\preprint{Phys. Review D}

\title{Scalable continuous gravitational wave detection in PTA data with non-parametric red noise suppression and optimal pulsar selection}

\author{Yi-Qian Qian}
\author{Yan Wang}%
\email{ywang12@hust.edu.cn}
\affiliation{%
  National Gravitation Laboratory, MOE Key Laboratory of Fundamental Physical Quantities Measurements, \\
  Department of Astronomy and School of Physics, Huazhong University of Science and Technology, Wuhan 430074, China
}%


\author{Soumya D. Mohanty}
\email{soumya.mohanty@utrgv.edu}
\affiliation{
  Department of Physics and Astronomy, The University of Texas Rio Grande Valley, Brownsville, TX 78520, United States of America \\
  Department of Physics, IIT Hyderabad, Kandai, Telangana-502284, India
}%

\author{Siyuan Chen}
\email{siyuan.chen@shao.ac.cn}
\affiliation{%
  State Key Laboratory of Radio Astronomy and Technology, Shanghai Astronomical Observatory, Chinese Academy of Sciences, Shanghai 200030, China
}%

\date{\today}

\begin{abstract}
  Bayesian methods for the detection of continuous gravitational waves (CGWs) in Pulsar Timing Array (PTA) data incur substantial computational costs that grow rapidly due to the number of noise and signal parameters characterizing the fitted model being proportional to the size of the PTA. This computational burden limits the scalability of these methods for large-scale PTAs comprising hundreds of pulsars anticipated from next-generation radio astronomy
  facilities. In this work, we introduce a computationally efficient frequentist method designed to circumvent this challenge. This is achieved by combining an adaptive spline fitting algorithm that non-parametrically suppresses red noise, thereby eliminating the need for complex noise modeling inherent to Bayesian methods, with a novel scheme for optimizing the subsets of pulsars included in the search. We quantify the performance of our method on a simulated dataset based on the NANOGrav 15-year data release and find that it achieves a performance comparable to that of Bayesian analysis: for a CGW signal with a signal-to-noise ratio of $\approx 10$, our method yields a relative characteristic strain error of 1.0\% and a frequency error of 0.072\% from the injected values by using the optimal pulsar selections, while the same errors are 1.7\% and 0.16\%, respectively, for the standard Bayesian analysis. At the same time, our analysis completes in less than 5 hours, in contrast to the 1-2 days required by Bayesian methods. This allows us to perform a rigorous study of our method using multiple data realizations and signal parameters, establishing it as an efficient and scalable tool for CGW searches with large-scale PTAs.

\end{abstract}

\maketitle

\section{Introduction \label{intro}}
Pulsar Timing Arrays (PTAs) are galactic-scale experiments designed to detect ultra-low-frequency gravitational waves (GWs) in the nanohertz band~\cite{sazhinOpportunitiesDetectingUltralong1978}. By precisely measuring the time of arrivals (ToAs) of radio pulses from an ensemble of millisecond pulsars, PTAs can detect the subtle spacetime perturbations caused by passing gravitational waves. The global effort for PTA-based GWs detection is spearheaded by several major collaborations, including the North American Nanohertz Observatory for Gravitational Waves (NANOGrav)~\cite{jenetNorthAmericanNanohertz2009, mclaughlinNorthAmericanNanohertz2013}, the Parkes Pulsar Timing Array (PPTA) in Australia~\cite{manchesterParkesPulsarTiming2013}, the European Pulsar Timing Array (EPTA)~\cite{kramerEuropeanPulsarTiming2013, babakEuropeanPulsarTiming2016}, the MeerKAT Pulsar Timing Array (MPTA)~\cite{bailesMeerTimeMeerKATKey2018}, the Indian Pulsar Timing Array (InPTA)~\cite{paulIndianPulsarTiming2019}, and the Chinese Pulsar Timing Array(CPTA)~\cite{leeProspectsGravitationalWave2016a}. These consortia also combine their data under the banner of the International Pulsar Timing Array (IPTA)~\cite{hobbsInternationalPulsarTiming2010} to enhance detection sensitivity.

The primary targets for PTA detection are GWs emanating from the cosmic population of supermassive black hole binaries (SMBHBs). These binaries are expected to form during galaxy mergers and the GWs emitted by them are expected to superpose to form a stochastic gravitational wave background (SGWB), while some of them may also be detected as individual, resolvable continuous gravitational waves (CGWs) from nearby or particularly massive systems~\cite{burke-spolaorAstrophysicsNanohertzGravitational2019}. The analysis of PTA data collected so far shows growing evidence for a nanohertz SGWB. NANOGrav~\cite{agazieNANOGrav15Yr2023b}, EPTA+InPTA~\cite{antoniadisSecondDataRelease2023c}, PPTA~\cite{reardonSearchIsotropicGravitationalwave2023}, MeerKAT~\cite{milesMeerKATPulsarTiming2024} and CPTA~\cite{xuSearchingNanoHertzStochastic2023} all reported various levels of evidence for an SGWB in their latest data releases and they are all consistent with each other~\cite{theinternationalpulsartimingarraycollaborationComparingRecentPTA2023}.

In parallel with the tentative discovery of an SGWB, the search for individual sources is a major focus for all leading PTA collaborations. Using their 15-year dataset, NANOGrav performed extensive searches for CGWs from dozens of candidate SMBHBs in nearby galaxies~\cite{agazieNANOGrav15Yr2023c}. Similarly, the EPTA, in its second data release, has conducted comprehensive searches for both targeted and all-sky CGW signals~\cite{antoniadisSecondDataRelease2024}. The most stringent constraints to date come from the IPTA, which leverages the combined sensitivity of multiple arrays to place upper limits on the GW emission from the most massive and nearest SMBHB candidates~\cite{falxaSearchingContinuousGravitational2023}. While no definitive detection of an individual CGW has been made yet, these efforts have successfully constrained the parameters of potential sources and continue to push the boundaries of detection.

The analysis of PTA data for these CGW signals is a significant challenge. Most of the published analysis of real PTA data adopt a Bayesian method for parameter estimation and model selection in either the search for SGWB or individual CGWs~\cite{ellisBayesianAnalysisPipeline2013,johnsonNANOGrav15yearGravitationalWave2023, becsyJointSearchIsolated2020}, and several software packages have been developed~\cite{ellisENTERPRISEEnhancedNumerical2020,luoPINTModernSoftware2021}. However, these techniques are often hindered by their high computational cost, as they require explicit and complex modeling of various noise components which will grow with the size of the PTA. Although, there are several techniques are introduced in the Bayesian method to accelerate the analysis~\cite{becsyFastBayesianAnalysis2022,becsyEfficientBayesianInference2024a,vallisneriRapidParameterEstimation2024a}, they are still unscalable for large-scale PTAs. Most importantly as datasets grow in size and complexity~\cite{wangPulsarTimingArray2017} with the advent of next-generation instruments, such as FAST~\cite{nanFIVEHUNDREDMETERAPERTURESPHERICAL2012} and Square Kilometer Array (SKA)~\cite{dewdneySquareKilometreArray2009,janssenGravitationalWaveAstronomy2015}, the need for more efficient analysis techniques that can scale to the large-scale PTAs becomes paramount.

In this paper, we introduce a novel and computationally efficient frequentist method for the detection of CGWs. Our method, referred to as the SHAPES+MaxAvPhase (SM) method, circumvents the challenge of explicit noise modeling by integrating two key algorithms. First, we utilize the Swarm Heuristics-based Adaptive and Penalized Estimation of Spline (\texttt{SHAPES}) algorithm~\cite{mohantyAdaptiveSplineFitting2021} to non-parametrically suppress intrinsic pulsar red noise~\cite{qianRedNoiseSuppression2025} via an adaptive spline fitting process. Following this noise suppression step, we employ the MaxAvPhase algorithm~\cite{wangCoherentMethodDetection2014, wangCoherentNetworkAnalysis2015,qianIterativeTimedomainMethod2022} to search for the CGW signal. This two-step approach provides an agile and powerful tool for large-scale CGW searches, reducing the analysis time from days, as required by Bayesian methods, to less than 5 hours.

In addition to the \texttt{SHAPES} algorithm, we introduce a novel pulsar selection scheme inspired by the `quality over quantity' concept proposed in~\cite{speriQualityQuantityOptimizing2023}. Strategic pulsar selection is essential for optimizing the search for CGWs because sensitivity to these deterministic signals is not uniformly distributed across a PTA. As demonstrated in~\cite{speriQualityQuantityOptimizing2023}, individual pulsars contribute varying amounts to the total SNR of a potential CGW source. This contribution depends on each pulsar's intrinsic noise properties and observing cadence. By ranking pulsars based on their SNR contributions for a CGW signal, allowing us to identify a small subset that captures the majority of the array's total sensitivity. Here, we numerically obtain the SNR for each pulsar for a given CGW source by injecting the CGW signal into the simulated timing residuals and calculating its SNR. We explored three different selection schemes and compared their performance against the SM and Bayesian methods. When employing the optimal strategy to select the most stable subset of pulsars, our method yielded a relative characteristic strain error of 1.0\% and a frequency error of 0.072\% from the injected values. In comparison, the Bayesian analysis resulted in a relative characteristic strain error of 1.7\% and a relative frequency error of 0.16\%. Besides, our method significantly enhances computational efficiency, completing the analysis in less than 5 hours, whereas the Bayesian methods typically require 1 to 2 days.

The rest of the paper is structured as follows: Sec.~\ref{data_model} describes the data model and the simulation of the data used in this study. In Sec.~\ref{methods}, we present the methodology of our method and the Bayesian method used in our comparison. In Sec.~\ref{results}, a comprehensive comparison between our method and the Bayesian method is presented, as well as robustness tests for the pulsar selection. Finally, in Sec.~\ref{conclusion}, we conclude with discussing the implications of our results.

\section{Data Model and Simulation \label{data_model}}
The timing residuals $\boldsymbol{\delta t}$ are a vector of data collected from various observation epochs. Within each epoch, there are multiple data collected from different observing frequency bands. Following the framework established in recent PTA studies~\cite{agazieNANOGrav15Yr2023c,falxaSearchingContinuousGravitational2023,antoniadisSecondDataRelease2024}, the data model is given by:
\begin{equation}
  \boldsymbol{\delta t} = \boldsymbol{M}\boldsymbol{\epsilon} + \boldsymbol{\delta n} + \boldsymbol{s}_{\text{SSE}} + \boldsymbol{s}_{\text{CGW}}\;.
\end{equation}
In this formulation, the errors, denoted by $\boldsymbol{\epsilon}$, in the best fit parameters of the timing model are included linearly, while $\boldsymbol{M}\boldsymbol{\epsilon}$ accounts for small perturbations from the best-fit timing model. $\boldsymbol{\delta n}$ is the noise which will be discussed in the Sec.~\ref{noise_model},  $\boldsymbol{s}_{\text{SSE}}$ mitigates potential systematic errors in the Solar System Ephemeris, and $\boldsymbol{s}_{\text{CGW}}$ is the deterministic signal from a single CGW source which is the focus of the present work.
\subsection{Signal Model}
A deterministic model for a CGW signal from a non-evolving, circular supermassive black hole binary (SMBHB) is given by~\cite{wangCoherentMethodDetection2014,wangCoherentNetworkAnalysis2015,qianIterativeTimedomainMethod2022,qianRedNoiseSuppression2025} and the timing residual $s_\mathrm{CGW}(t; \hat{\Omega})$ induced by the CGW signal is given by:
\begin{equation}
  s_\mathrm{CGW}(t, \hat{\Omega})=\sum_A F^A(\hat{\Omega})\left[s_A(t)-s_A\left(t-\tau_a\right)\right] \;,\\
\end{equation}
with
\begin{equation}
  \begin{aligned}
    s_{+}(t)= & \frac{\mathcal{M}^{5 / 3}}{d_L \omega(t)^{1 / 3}}\left[-\sin [2 \Phi(t)]\left(1+\cos ^2 \iota\right) \cos 2 \psi\right. \\
    & -2 \cos [2 \Phi(t)] \cos \iota \sin 2 \psi]\;, \\
    s_{\times}(t)= & \frac{\mathcal{M}^{5 / 3}}{d_L \omega(t)^{1 / 3}}\left[-\sin [2 \Phi(t)]\left(1+\cos ^2 \iota\right) \cos 2 \psi\right. \\
    & +2 \cos [2 \Phi(t)] \cos \iota \sin 2 \psi]\;. \\
  \end{aligned}
\end{equation}
Here, $A$ denotes the $+$ and $\times$ polarization of the CGWs, $\mathcal{M}$ is the chirp mass, $d_L$ is the luminosity distance, $\omega(t)$ is the orbital frequency, $\Phi(t)$ is the phase, $\iota$ is the inclination angle, and $\psi$ is the polarization angle. $F^A$ are called the antenna pattern functions~\cite{sesanaMeasuringParametersMassive2010,babakResolvingMultipleSupermassive2012} which can be expressed as:
\begin{equation}
  \begin{aligned}
    F^{+}(\hat{\Omega}) & =\frac{1}{2} \frac{(\hat{m} \cdot \hat{p})^2-(\hat{n} \cdot \hat{p})^2}{1+\hat{\Omega} \cdot \hat{p}} \;,\\
    F^{\times}(\hat{\Omega}) & =\frac{(\hat{m} \cdot \hat{p})(\hat{n} \cdot \hat{p})}{1+\hat{\Omega} \cdot \hat{p}} \;,
  \end{aligned}
\end{equation}
in which, $\hat{\Omega}$ is the direction of the GW propagation, $\hat{p}$ is the direction of the pulsar, $\hat{m}$ and $\hat{n}$ are two unit vectors defined as:
\begin{equation}
  \begin{aligned}
    & \hat{m}=-\sin \phi \hat{x}+\cos \phi \hat{y}\;, \\
    & \hat{n}=-\cos \theta \cos \phi \hat{x}-\cos \theta \sin \phi \hat{y}+\sin \theta \hat{z}\;,
  \end{aligned}
\end{equation}
where, $(\theta, \phi)$ are the polar coordinates of the source position on the sky.

The signal-to-noise ratio (SNR) is a metric used to characterize the strength of a CGW signal. For a full PTA, the network SNR can be defined as follows:
\begin{equation}\label{eq:snr_formula}
  \begin{aligned}
    \mathrm{SNR} & =\left[\sum_{I=1}^{N_p}\left(\rho^I\right)^2\right]^{1 / 2} \;,\\
    \rho^I & =\left(\boldsymbol{s}_\mathrm{CGW}^I\right)^T \cdot \boldsymbol{C}^{-1} \cdot \boldsymbol{s}_\mathrm{CGW}^I \;.
  \end{aligned}
\end{equation}
Here, $N_p$ is the number of pulsars in the PTA, $\rho^I$ is the SNR for the $I$-th pulsar, $\boldsymbol{C}$ is the noise covariance matrix, and $\boldsymbol{s}_\mathrm{CGW}^I$ is the timing residual induced by the CGW signal for the $I$-th pulsar.

\subsection{Noise Model}
\label{noise_model}
A phenomenological noise model $\boldsymbol{\delta n}$ is commonly employed for both time-uncorrelated and time-correlated noise components in the data, which can be expressed as:
\begin{equation}
  \boldsymbol{\delta n} = \boldsymbol{n}_{\text{WN}} + \boldsymbol{n}_{\text{IRN}} + \boldsymbol{n}_{\text{CURN}} \;.
\end{equation}
Here, $\boldsymbol{n}_{\text{WN}}$ is a white noise component modeling uncorrelated measurement errors such as radiometer noise and pulse jitter. The white noise $\boldsymbol{n}_{\mathrm{WN}}$ is defined through its covariance matrix, $\boldsymbol{C}$, which is formulated as $\boldsymbol{C} = \Phi^2(\boldsymbol{\sigma_{\text{S/N}}}^2 + \Theta^2) + J^2\boldsymbol{U}$. In this expression, $\Phi$ (EFAC) is a scaling factor for the formal ToA uncertainties, denoted by $\boldsymbol{\sigma_{\text{S/N}}}$, while $\Theta$ (EQUAD) is an additional noise term added in quadrature. While ``white noise'' usually refers to uncorrelated errors, jitter noise in narrow-band timing is frequency-correlated within an observing epoch. It is modeled using $J$ (ECORR), where the block-diagonal matrix $\boldsymbol{U}$ governs its influence. This matrix assigns a value of 1 to ToAs from the same epoch across different observing frequency bands or receivers, and 0 otherwise~\cite{agazieNANOGrav15Yr2023a}. EFAC and EQUAD set the variance for individual ToAs while ECORR introduces the off-diagonal, block-diagonal correlations. $\boldsymbol{n}_{\text{IRN}}$ represents an achromatic, time-correlated intrinsic red noise (IRN) process intrinsic to each pulsar. For a specific pulsar $i$, it can be characterized by two parameters: the amplitude $A_{\mathrm{IRN}}^i$ and the spectral index $\gamma_{\mathrm{IRN}}^i$; $\boldsymbol{n}_{\text{CRN}}$ is so-called common (spatially) uncorrelated red noise (CURN) process which has the same power spectral density (PSD) across all pulsars in the array, both IRN and CURN can be modeled by Eq.~(\ref{eq:psd_rn}) with the amplitude $A_{\mathrm{xRN}}^{(i)}$ and spectral index $\gamma_{\mathrm{xRN}}^{(i)}$:
\begin{equation}\label{eq:psd_rn}
  P_{\mathrm{xRN}}^{(i)}(f)=\frac{A_{\mathrm{xRN}}^{(i)2}}{12 \pi^2} f_\mathrm{yr}^{-3}\left(\frac{f}{f_\mathrm{yr}}\right)^{-\gamma_{\mathrm{xRN}}^{(i)}} \,.
\end{equation}
Here, $A_\mathrm{xRN}^{(i)} = A_\mathrm{IRN}^i$, $\gamma_\mathrm{xRN}^{(i)} = \gamma_\mathrm{IRN}^i$ for IRN,  $A_\mathrm{xRN}^{(i)} = A_\mathrm{CURN}$, $\gamma_\mathrm{xRN}^{(i)} = \gamma_\mathrm{CURN}$ for CURN, and $f_\mathrm{yr} = 1~\mathrm{yr}^{-1}$ is the commonly adopted reference frequency.

In this study, we ignore the contribution from chromatic noise due to time-varying dispersion measure (DM) effects. We also ignore the ECORR term in the noise model, since we only use a single frequency in our data simulations which will be described in details in the next section. Additionally, the DM, EFAC, and EQUAD parameters are fixed at certain values.

\subsection{Data Simulation}
The data used in the subsequent study are simulated based on the NANOGrav 15-year dataset (NG15)~\cite{agazieNANOGrav15Yr2023} using the \texttt{PINT}~\cite{luoPINTModernSoftware2021} and the \texttt{ENTERPRISE}~\cite{ellisENTERPRISEEnhancedNumerical2020} software packages. The NG15 is a collection of pulsar timing observations that spanned over 15 years for 68 millisecond pulsars. This is a significant expansion from the 47 pulsars in the prior 12.5-year date release~\cite{alamNANOGrav12Yr2020a}. These data are obtained using the Arecibo Observatory, the Green Bank Telescope (GBT), and the Very Large Array (VLA), with the longest individual pulsar timing baseline extending to 16 years. The observations were conducted on cadences ranging from days to months across a wide range of radio frequencies. The dataset also includes the models for stochastic noise components, such as IRN, which is considered a critical component, especially for pulsars with long timing baselines. In this dataset, 23 of the observed pulsars are identified with significant levels of IRN.

\begin{figure}[htbp]
  \centering
  \includegraphics[width=0.5\textwidth]{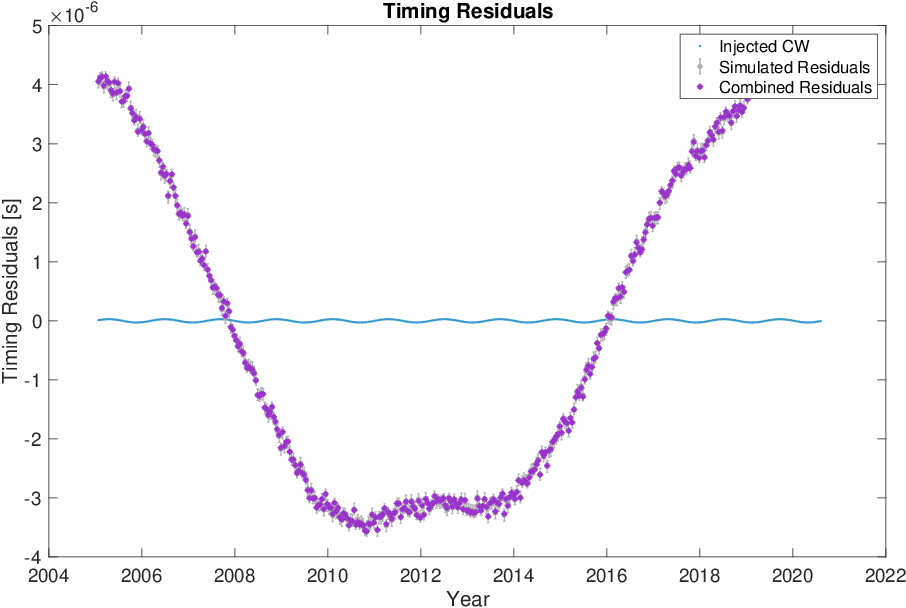}
  \caption{The simulated timing residuals for the pulsar J0030+0451. Blue dots represent the injected timing residuals induced by the CGW signal, the gray dots show the simulated noises using the \texttt{PINT}, and the purple dots are the sum of the two.}
  \label{fig:J0030+0451_timing_residual}
\end{figure}

For each pulsar's simulated timing residuals, we use the same start and end observation time with the same number of observations as the NG15 data, but with a biweekly observation cadence. We also use the same timing model and noise parameters as the NG15 data, but fixing the timing precision of each time of arrival (ToA) to 100 ns. After getting the simulated timing residuals, we inject a CGW signal with an SNR of approximately 10 and a frequency of 26 nHz into the dataset. Fig.~\ref{fig:J0030+0451_timing_residual} shows the simulated noise component in the timing residuals, the injected signal, and their combination for the pulsar J0030+0451 in the NG15 that presents significant IRN.

\section{METHODS \label{methods}}
In this section, we first introduce three different pulsar selection schemes and a non-parametric red noise suppression method. Next, we briefly explain the setup for our frequentist method and the Bayesian method to be compared in Sec.~\ref{results}.

\subsection{Pulsar Optimization \label{pulsar_optimization}}

It was demonstrated in \cite{speriQualityQuantityOptimizing2023} that a large portion of a PTA's sensitivity is often concentrated within a smaller, optimal subset of pulsars. While a standard likelihood-based analysis can judiciously integrate the data from all pulsars when their noise models are known, this does not need be the case for real data where the characteristics of noise need to be estimated from real data. This implies that the inclusion of numerous lower-quality pulsars can yield diminishing returns while substantially increasing the computational burden of the analysis. This challenge is expected to intensify with the discovery of many new pulsars by next-generation radio facilities.

\begin{figure*}[htbp]
  \centering
  \subfigure[C-SNR-90 scheme]{
    \includegraphics[width=0.28\textwidth]{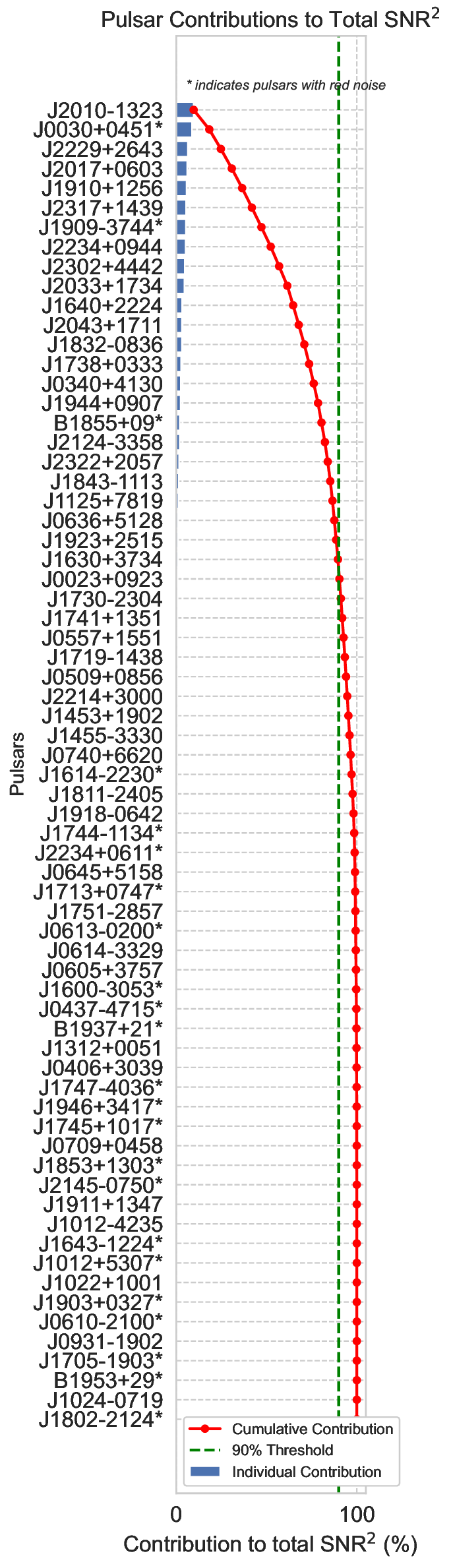}
    \label{fig:pulsar_contributions_90}
  }
  \hfill
  \subfigure[C-ASNR-90 scheme]{
    \includegraphics[width=0.38\textwidth]{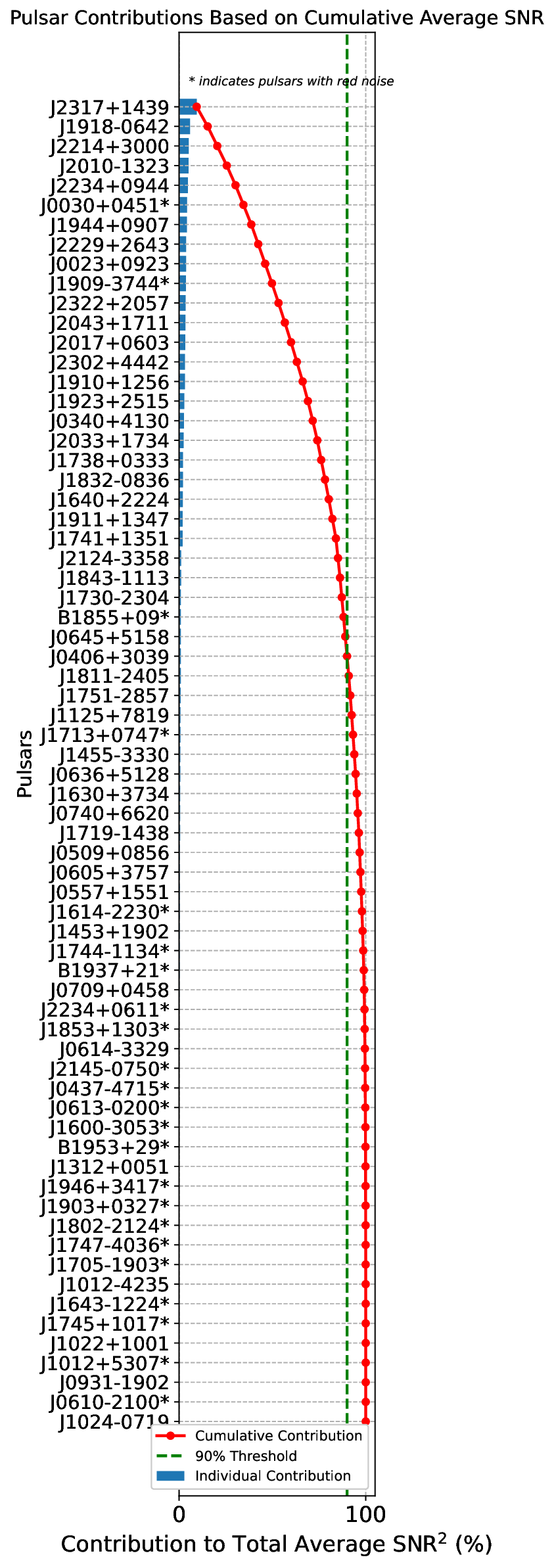}
    \label{fig:avg_snr_contributions}
  }
  \hfill
  \subfigure[P-60 scheme]{
    \includegraphics[width=0.28\textwidth]{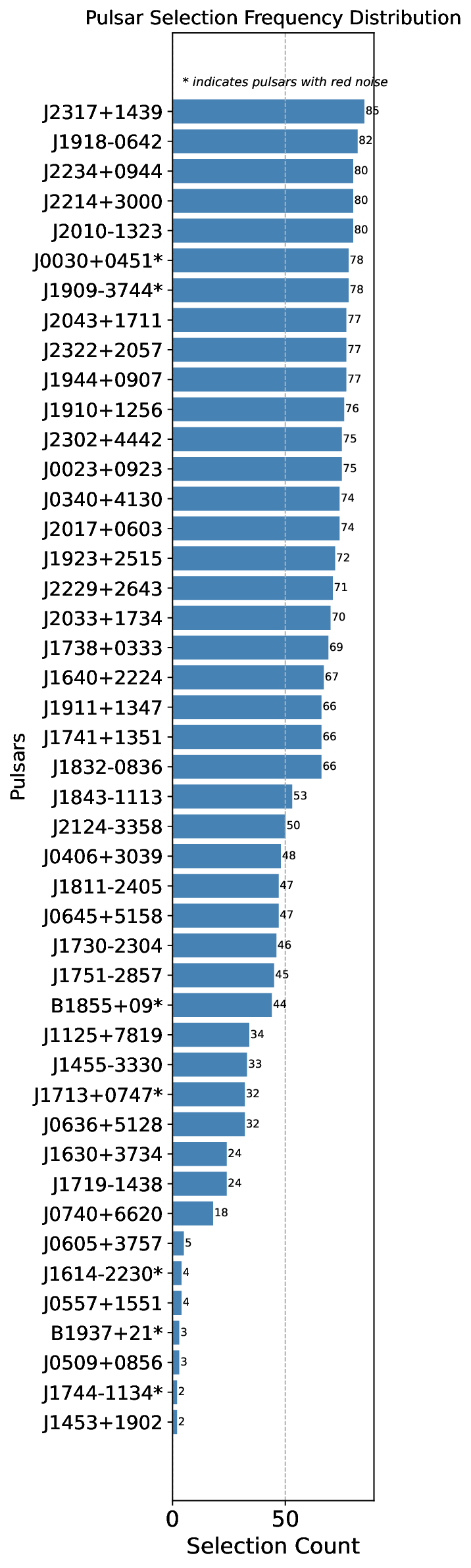}
    \label{fig:hf60U_selection_frequency}
  }
  \caption{Comparison of the pulsar subsets obtained using the three selection schemes described in Sec.~\ref{pulsar_optimization}. (a) C-SNR-90 scheme: The histogram displays the distribution of each pulsar's SNR contribution. The red line indicates the pulsars' cumulative contribution, and the green dashed line represents the 90\% threshold of the total SNR. (b) C-ASNR-90 scheme: The histogram displays the distribution of each pulsar's average SNR contribution. The red line indicates the cumulative contribution of the pulsars, while the green dashed line represents the 90\% threshold of the total average SNR. (c) P-60 scheme: The number `60' is a manually chosen threshold, identifying pulsars selected more than 60 times out of 100 frequency realizations. The histogram displays how frequently each pulsar is selected. Pulsars with intrinsic red noise are indicated by an asterisk after their names.}
  \label{fig:pulsar_selection_schemes}
\end{figure*}

In~\cite{speriQualityQuantityOptimizing2023}, the authors proposed an approach to optimize the pulsar selection according to each pulsar's response to a CGW signal. The core of this technique is to quantify each pulsar's contribution via averaging its contribution to the network SNR over the unknown source parameters, such as sky location, inclination, and polarization angle. This can be achieved either numerically by injecting and averaging over a large set of simulated signals, or through a direct analytical formula~\cite{speriQualityQuantityOptimizing2023}. Recognizing that the pulsar ranking is frequency-dependent, the final selection is performed by merging ranked lists generated across multiple frequency bins. In this work, we build upon this basic idea by exploring three distinct pulsar selection schemes:

\begin{enumerate}
  \item \textbf{Cumulative SNR-X (C-SNR-X) scheme:}
    For a specific CGW source, the SNR contribution for each pulsar is calculated using Eq.~\ref{eq:snr_formula}. We then sort these contributions in descending order and select the pulsars whose cumulative contribution reaches X\% of the total SNR. Setting X to 90, we get the C-SNR-90 scheme, as illustrated in Fig.~\ref{fig:pulsar_contributions_90}.

  \item \textbf{Cumulative Average SNR-X (C-ASNR-X) scheme:}
    For a given source sky location, 100 values of the signal frequency are randomly selected from a uniform grid of values in the range 20~nHz to 30~nHz. For each pulsar, the SNR is averaged across this set of signal frequencies. We then sort these average contributions in descending order and select those pulsars whose contribution to the total SNR reaches X\%. Setting X to 90, we get the C-ASNR-90 scheme.

  \item \textbf{Persistence (P-Y) scheme:}
    Similar to the C-ASNR-X scheme, we compute the SNR values for each pulsar at 100 uniformly spaced signal frequencies. Then, for each frequency value, we sort the pulsars in descending order of SNRs and note the pulsars that contribute X\% of the total SNR. The subset of pulsars selected in this manner changes with the signal frequency values. Only the pulsars that are selected Y times (out of 100) are retained. Setting X to 90 and Y to 60, we get the P-60 scheme.
\end{enumerate}

Table~\ref{tab:pulsar_difference} presents the pulsars selected under these three schemes for a GW source located at location A ((RA, DEC) = (5.96, 0.21)). Under the C-ASNR-90, C-SNR-90, and P-60 schemes, the numbers of selected pulsars are 30, 25, and 23, respectively. Thus, the pulsar subsets picked by the different schemes are not identical. The intersection of the subsets has 19 pulsars. The C-ASNR-90 scheme has the largest complement of the intersection set with 11 pulsars, while the P-60 scheme has the smallest with 4 pulsars. Both C-ASNR-90 scheme and C-SNR-90 scheme select the pulsar B1855+09, which has significant IRN. The search results of these two schemes are very similar across the Bayesian and our SM methods, we will discuss it in details in Sec.~\ref{results}.
\begin{table}[h!]
  \centering
  \caption{Pulsar subsets obtained using the three selection schemes described in Sec.~\ref{pulsar_optimization}. The upper portion shows the subset of pulsars that are in common to the three selection schemes. The lower portion gives the lists of the pulsars that are unique to each scheme. The scheme names are shown at the bottom. Asterisk (*) indicates pulsars with intrinsic red noise.}
  \label{tab:pulsar_difference}
  \begin{tabular}{|l|l|l|}
    \hline
    \multicolumn{3}{|c|}{Common PSRs} \\
    \hline
    \multicolumn{3}{|c|}{J0023+0923} \\
    \multicolumn{3}{|c|}{J0030+0451*} \\
    \multicolumn{3}{|c|}{J0340+4130} \\
    \multicolumn{3}{|c|}{J1640+2224} \\
    \multicolumn{3}{|c|}{J1738+0333} \\
    \multicolumn{3}{|c|}{J1832-0836} \\
    \multicolumn{3}{|c|}{J1909-3744*} \\
    \multicolumn{3}{|c|}{J1910+1256} \\
    \multicolumn{3}{|c|}{J1923+2515} \\
    \multicolumn{3}{|c|}{J1944+0907} \\
    \multicolumn{3}{|c|}{J2010-1323} \\
    \multicolumn{3}{|c|}{J2017+0603} \\
    \multicolumn{3}{|c|}{J2033+1734} \\
    \multicolumn{3}{|c|}{J2043+1711} \\
    \multicolumn{3}{|c|}{J2229+2643} \\
    \multicolumn{3}{|c|}{J2234+0944} \\
    \multicolumn{3}{|c|}{J2302+4442} \\
    \multicolumn{3}{|c|}{J2317+1439} \\
    \multicolumn{3}{|c|}{J2322+2057} \\
    \hline
    \multicolumn{3}{|c|}{Different PSRs} \\
    \hline
    B1855+09* & J1741+1351 & B1855+09* \\
    J0636+5128 & J1911+1347 & J0406+3039 \\
    J1125+7819 & J1918-0642 & J0645+5158 \\
    J1630+3734 & J2214+3000 & J1730-2304 \\
    J1843-1113 & & J1741+1351 \\
    J2124-3358 & & J1811-2405 \\
    & & J1843-1113 \\
    & & J1911+1347 \\
    & & J1918-0642 \\
    & & J2124-3358 \\
    & & J2214+3000 \\
    \hline
    \multicolumn{1}{|c|}{\textbf{C-SNR-90}} & \multicolumn{1}{c|}{\textbf{P-60}} & \multicolumn{1}{c|}{\textbf{C-ASNR-90}} \\
    \hline
  \end{tabular}
\end{table}

\subsection{Red Noise mitigation \label{shapes_maxavphase_framework}}
We use the SHAPES-based red noise mitigation method introduced in~\cite{qianRedNoiseSuppression2025} to reduce the red noise in the timing residuals. This method is based on an adaptive spline fitting approach and the algorithm~\cite{mohantyAdaptiveSplineFitting2021} implemented in \texttt{MATLAB}~\cite{Mohanty_SHAPES_2024}. It is described in detail in~\cite{mohantyAdaptiveSplineFitting2021} and has been successfully applied to subtract glitches in LIGO data~\cite{mohantyGlitchSubtractionGravitational2023a}. The regression model used in \texttt{SHAPES} posits that the observed data vector, $\boldsymbol{y}$, is the sum of a trend\footnote{Any process which deviates from white noise.}, $\boldsymbol{s}(\Gamma)$, and additive white noise, $\boldsymbol{\epsilon}$, represented as $\boldsymbol{y} = \boldsymbol{s}(\Gamma) + \boldsymbol{\epsilon}$. Its core principle involves modeling the continuous signal $s(t; \Gamma)$ as a spline, composed of piecewise cubic polynomials expressed as a linear combination of B-spline basis functions~\cite{de1978practical}, $B_{j,4}(t; \boldsymbol{\tau})$:
\begin{equation}
  s(t; \Gamma = \{\boldsymbol{\alpha}, \boldsymbol{\tau}\}) = \sum_{j} \alpha_j B_{j,4}(t; \boldsymbol{\tau})\;,\\
\end{equation}
where $\boldsymbol{\alpha}$ are the B-spline coefficients and $\boldsymbol{\tau}$ is the vector of knot locations. The best-fit parameters are ultimately derived by minimizing a penalized least-squares function:
\begin{equation}
  \sum_{i=1}^{N} (y_i - s_i(\Gamma))^2 + \lambda \sum_{j=0}^{P-5} \alpha_j^2 \;.
\end{equation}
The user-specified penalty gain, $\lambda$, governs the smoothness of the resulting spline estimate. \texttt{SHAPES} utilizes the Particle Swarm Optimization (PSO)~\cite{kennedyParticleSwarmOptimization1995, mohantyParticleSwarmOptimization2018} metaheuristic to find the optimal knot locations $\boldsymbol{\tau}$ for a given knot numbers $P$, allowing multiple knots to coalesce to capture point discontinuities, while the Akaike Information Criterion (AIC)~\cite{AkaikeInformationTheory1998} is used for model selection to identify the optimal $P$, balancing goodness-of-fit with model complexity. The current implementation of \texttt{SHAPES} is limited to regularly sampled data, which is the main reason we have chose to use uniformly sampled data in our simulations. However, the underlying de Boor recursion relations for obtaining B-splines are defined for any type of sampling, and a version of \texttt{SHAPES} for non-uniformly sampled data is planned for future work.
\begin{figure*}[htbp]
  \centering
  \subfigure[\texttt{SHAPES} detrended timing residuals for J0030+0451]{
    \includegraphics[width=0.8\textwidth]{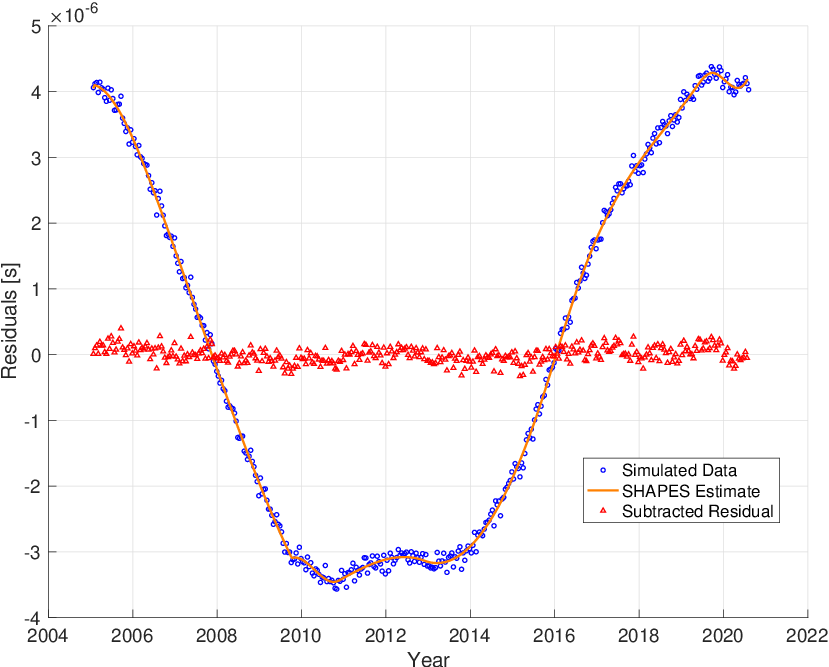}
    \label{fig:J0030+0451_comb_detrend_lp_Ns1}
  }
  \vfill
  \subfigure[Lomb-Scargle periodogram for J0030+0451.]{
    \includegraphics[width=0.8\textwidth]{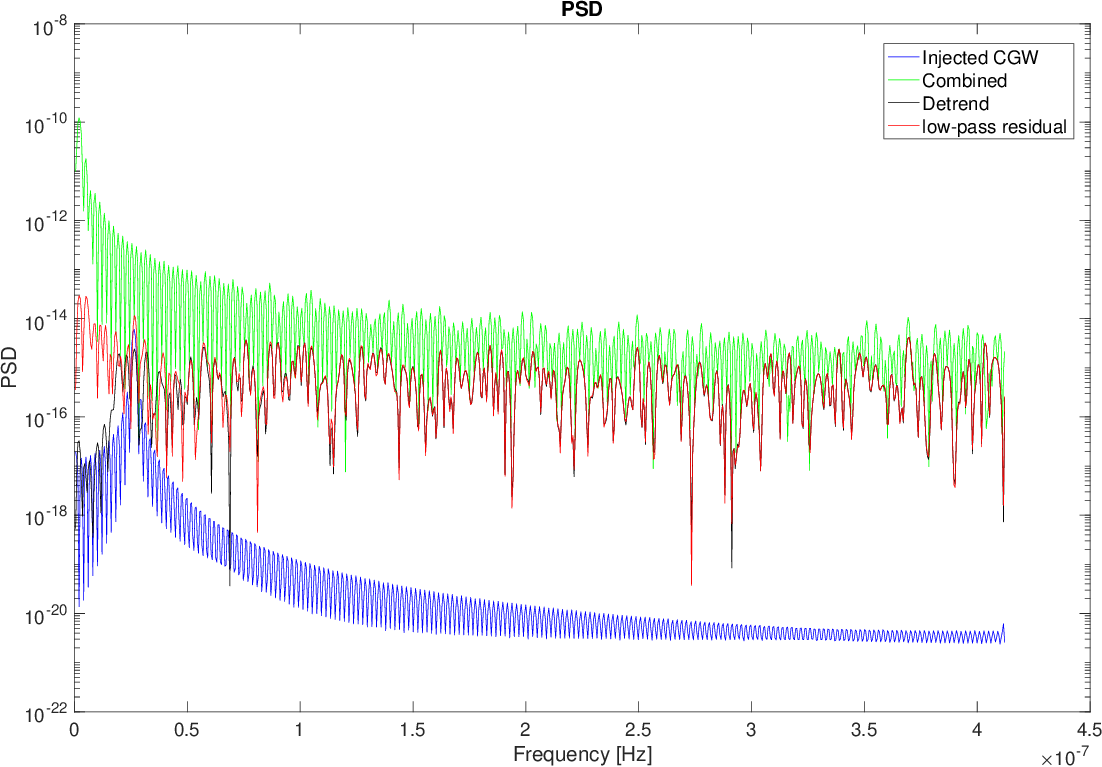}
    \label{fig:J0030+0451_LSP_lp_r8}
  }
  \caption{\texttt{SHAPES} detrended timing residuals and Lomb-Scargle periodograms for PSR J0030+0451. The top panel shows detrended timing residuals, and the bottom panel shows the Lomb-Scargle periodograms for the injected CGW signal (blue), the data with injected CGW signal (green), the \texttt{SHAPES} detrended timing residuals (black) and the \texttt{SHAPES} detrended timing residuals with a low-pass filter (red).}
  \label{fig:J0030+0451_SHAPES}
\end{figure*}
Fig.~\ref{fig:J0030+0451_comb_detrend_lp_Ns1} illustrates the application of the \texttt{SHAPES} method. The process begins by fitting the timing residuals, which yields the \texttt{SHAPES} estimates. Next, to suppress the low frequency content and preserve the power from the signal, a finite impulse response (FIR) low-pass filter is applied to the \texttt{SHAPES} estimates. The filter's order, $N^I_\mathrm{order}$, for pulsar $I$ is determined by the number of data points ($N^I_{\mathrm{sample}}$) for that pulsar, specifically $N^I_\mathrm{order}=\texttt{floor}(N^I_\mathrm{sample}/R) - 1$. Here, the hyperparameter $R$ is empirically set to 8 for best performance. For simplicity, the cutoff frequency of the low-pass filter is set to 20 nHz for all pulsars, although this too can be made an adjustable parameter in future versions of the method. The \texttt{SHAPES} with a low-pass filter suppresses low-frequency content while preserves the power of the injected CGW signal, as shown in Fig.~\ref{fig:J0030+0451_LSP_lp_r8}. This is also demonstrated in Fig.~5 of~\cite{qianRedNoiseSuppression2025}for PSR J1909-3744, and more details about this pulsar can be found therein.

\subsection{CGW Search: frequentist method \label{frequentist_approach}}
Following detrending of the red noise for all pulsars, we employ the MaxAvPhase algorithm implemented in the \texttt{SAPTARISHI} code~\cite{wangCoherentNetworkAnalysis2015,qianIterativeTimedomainMethod2022} to search for the CGW signal as the second step of the SM method. Here, we decompose the source parameter $\boldsymbol{\lambda}$ into two parts: seven intrinsic parameters ($\boldsymbol{\lambda}_i$) and $N_p$ extrinsic parameters, which represents the pulsar phases ($\boldsymbol{\phi}_I$), the definition of $\boldsymbol{\phi}_I$ can be found in~\cite{wangCoherentMethodDetection2014}, and the pulsar distances we used in this analysis are from NANOGrav official CW analysis Github repository\footnote{https://github.com/nanograv/15yr\_cw\_analysis}. The intrinsic parameters include GW frequency, sky location (RA, DEC), GW amplitude, inclination angle, GW polarization angle, and orbital initial phase. The MaxAvPhase method estimates these intrinsic parameters by maximizing the marginalized log-likelihood function, which involves integrating over the extrinsic parameters semi-analytically. This estimation process is guided by the Marginalized-Maximized Likelihood Ratio Test (MMLRT) statistic, defined as~\cite{qianIterativeTimedomainMethod2022}:
\begin{equation}
  \operatorname{MMLRT}(\boldsymbol{\mathcal{Y}})=\max _{\boldsymbol{\lambda}_i} \ln \left(\operatorname{marg}_{\boldsymbol{\phi}_I} \Lambda(\boldsymbol{\mathcal{Y}} ; \boldsymbol{\lambda})\right)\;.\\
\end{equation}
Here, $\Lambda(\boldsymbol{\mathcal{Y}}; \boldsymbol{\lambda})$ is the network log-likelihood function for a PTA, and $\boldsymbol{\mathcal{Y}}$ is the timing residuals for all pulsars given whole parameters $\boldsymbol{\lambda}$. Subsequently, we use the estimated intrinsic parameters found by AvPhase to get the point estimates of the extrinsic parameters $\boldsymbol{\phi}_I$ using Maxphase. This method can perform a robust intrinsic parameter estimation while maintaining the full set of parameters which are needed to reconstruct the signal waveform.

\subsection{CGW Search: Bayesian method \label{bayesian_framework}}
In this work, we follow the standard Bayesian search setup for a CGW signal used in the NG15 dataset~\cite{agazieNANOGrav15Yr2023c} and IPTA DR2~\cite{falxaSearchingContinuousGravitational2023}.
In contrast to the MaxAvPhase~\cite{wangCoherentNetworkAnalysis2015,qianIterativeTimedomainMethod2022}, the Bayesian method usually considers the pulsar terms as part of the noise and ignore them in the data model to avoid the large dimensionality of the parameter space (2 additional parameters per pulsar for frequency and phase of pulsar term).
We use the \texttt{ENTERPRISE} software package~\cite{ellisENTERPRISEEnhancedNumerical2020} and PTMCMCSampler~\cite{justin_ellis_2017_1037579} to perform the MCMC search for $10^7$ steps with a thinning factor of 10, resulting $10^6$ samples. The first about $25\%$ of the samples are discarded as burn-in. Further details are provided in Appendix~\ref{app:bayesian_setup}. Table~\ref{table:bayesian_parameters} lists the priors of the parameters used in the Bayesian model. Due to the computational power limit, we only perform the Bayesian search for one data realization across the three different pulsar selection schemes, and the details are discussed in Sec.~\ref{bayesian_and_sm_search_results}.

\begin{table}[htbp]
  \centering
  \caption{Bayesian search parameters and priors}
  \label{table:bayesian_parameters}
  \begin{tabular}{ll}
    \toprule
    \textbf{Parameter} & \textbf{Prior} \\
    \midrule
    $\Phi$& 1\\
    $\Theta$ & $10^{-20}$\\
    $\log_{10} A_{\mathrm{IRN}}$ & $\mathcal{U}(-20, -10)$ \\
    $\log_{10} A_{\mathrm{CURN}}$ & $\mathcal{U}(-20, -10)$\\
    $\gamma_{\mathrm{IRN}}$ & $\mathcal{U}(0,7)$\\
    $\gamma_{\mathrm{CURN}}$ & $\mathcal{U}(0, 7)$\\
    $\log_{10} h$ & $\mathcal{U}(-18, -11)$ \\
    $f$ [Hz] & $\mathcal{U}(10^{-9}, 10^{-7})$ \\
    $\cos \iota$ & $\mathcal{U}(-1, 1)$ \\
    $\psi$ [rad] & $\mathcal{U}(0, \pi)$ \\
    $\phi_0$ [rad] & $\mathcal{U}(0, 2\pi)$ \\
    $\cos \theta$ & $\mathcal{U}(-1, 1)$ \\
    $\phi$ [rad] & $\mathcal{U}(0, 2\pi)$ \\
    \bottomrule
  \end{tabular}
\end{table}

\section{Results \label{results}}

In this section, we present a comprehensive evaluation of the proposed method. In Sec.~\ref{bayesian_and_sm_search_results}, we compare the detection performance and parameter estimation accuracy of our SM method against the standard Bayesian method across different pulsar subsets, using a single data realization. Subsequently, in Sec.~\ref{robustness_check_for_3_sets_of_optimized_pulsars}, we assess the statistical robustness of the three pulsar selection schemes by applying the SM method to 100 independent noise realizations and evaluating the stability of the results against variations in the source sky location.

\subsection{Comparison with Bayesian search \label{bayesian_and_sm_search_results}}
Due to the high computational cost of the Bayesian search, we are constrained by available resources to a comparison of the Bayesian and SM methods on just one data realization. However, this is sufficient for the purpose of demonstrating that the results are comparable. We compare their performance for the three subsets of pulsars selected by the C-SNR-90, C-ASNR-90, and P-60 schemes (c.f., Sec.~\ref{pulsar_optimization}). The results for the full PTA, where both methods perform poorly due to the inclusion of pulsars with strong intrinsic red noise, are presented in Appendix~\ref{app:full_pta_results}.

Fig.~\ref{fig:bayesian_avg}, Fig.~\ref{fig:bayesian_hf60U}, and Fig.~\ref{fig:bayesian_selected_90} show the corner plots for the Bayesian search results for the three optimized pulsar subsets. The Bayesian search succeeds in finding posterior distributions with well defined shapes for certain parameters for all three subsets, demonstrating the effectiveness of the pulsar selection schemes.

Table~\ref{tab:pulsar_error_comparison} presents a quantitative comparison of parameter estimation errors across the three algorithms—MaxAvPhase (without using \texttt{SHAPES}), SM, and Bayesian—applied to the three optimized pulsar subsets. The performance is evaluated based on relative errors in SNR ($\delta \text{SNR}/\text{SNR}$), characteristic strain ($\delta \log_{10} h / \log_{10} h$), and frequency ($\delta f / f$). Note that due to computational constraints, SNR errors for the Bayesian method are not reported.

The MaxAvPhase-only method consistently exhibits the poorest performance, yielding substantial relative errors across all metrics, with SNR errors exceeding 279\% and frequency errors staying near 80\% regardless of the dataset. These results illustrate the severe impact of unmitigated red noise on CGW searches. In contrast, the SM and Bayesian methods demonstrate significantly higher precision. The SM algorithm, by leveraging \texttt{SHAPES} for non-parametric red noise suppression, achieves SNR errors between 7.84\% and 15.69\% and frequency errors between 0.072\% and 0.49\% for the optimized pulsar subsets.

The SM algorithm performs particularly well on the P-60 and C-ASNR-90 sets, achieving frequency errors as low as 0.07\% and 0.35\%. Notably, when applied to the pulsars selected by the P-60 scheme, the accuracy achieved by the SM method for SNR estimation is comparable to the Bayesian method while offering a superior frequency estimate for this data realization. In contrast, both the SM and Bayesian methods perform significantly worse when the full PTA is used, as pulsars with strong intrinsic red noise effectively mask the true signal; a detailed analysis of this effect is provided in Appendix~\ref{app:full_pta_results}. This finding underscores the importance of pulsar quality in CGW searches.
\begin{table}[htbp]
  \centering
  \caption{Comparison of relative errors of parameter estimation for three optimized pulsar sets. The full PTA results are presented separately in Appendix~\ref{app:full_pta_results}.}
  \label{tab:pulsar_error_comparison}
  \begin{tabular}{cccccc}
    \toprule
    \textbf{Pulsar Sets} & \textbf{Methods} & $\frac{\delta \mathrm{SNR}}{\mathrm{SNR}}^1$ & $\frac{\delta \log_{10} h}{\log_{10} h}$ & $\frac{\delta f}{f}$ \\
    \midrule
    \multirow{3}{*}{\texttt{C-SNR-90}} & MaxAvPhase & 281.7\% & 21.58\% & 80.83\% \\
    & SM & 15.69\% & 0.8983\% & 0.4884\% \\
    & Bayesian & -- & 3.163\% & 0.1627\% \\
    \midrule
    \multirow{3}{*}{\texttt{C-ASNR-90}} & MaxAvPhase & 279.4\% & 23.79\% & 80.85\% \\
    & SM & 13.62\% & 2.223\% & 0.3499\% \\
    & Bayesian & -- & 1.676\% & 0.1627\% \\
    \midrule
    \multirow{3}{*}{\texttt{P-60}} & MaxAvPhase & 287.5\% & 18.35\% & 80.85\% \\
    & SM & 7.840\% & 1.005\% & 0.07212\% \\
    & Bayesian & -- & 6.271\% & 0.1627\% \\
    \bottomrule
  \end{tabular}
  \par
  \raggedright 
  \textsuperscript{1}\footnotesize{Relative SNR errors for the Bayesian method are not reported due to computational constraints.}
\end{table}

\begin{figure}[htbp]
  \centering
  \includegraphics[width=0.48\textwidth,height=0.45\textheight,keepaspectratio]{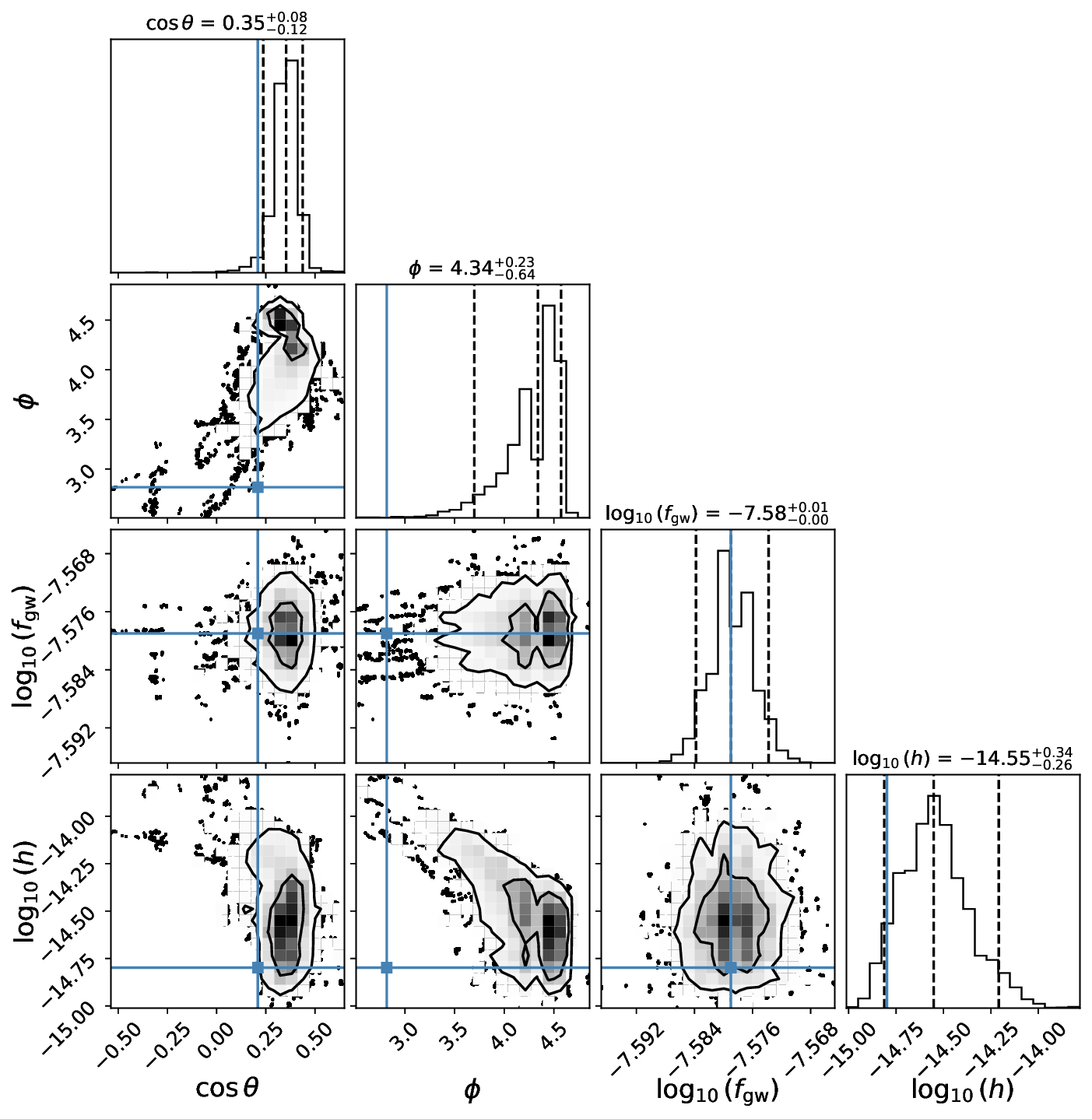}
  \caption{Bayesian search results for C-ASNR-90 optimized pulsar selection scheme. The corner plot shows the 1D marginal histograms with 5\%, 50\%, and 95\% quantiles, and the 2D joint marginal distributions for the CGW parameters, including frequency, sky location, and characteristic strain. The blue solid line represents the injected values of the CGW parameters. }
  \label{fig:bayesian_avg}
\end{figure}

\begin{figure}[htbp]
  \centering
  \includegraphics[width=0.48\textwidth,height=0.45\textheight,keepaspectratio]{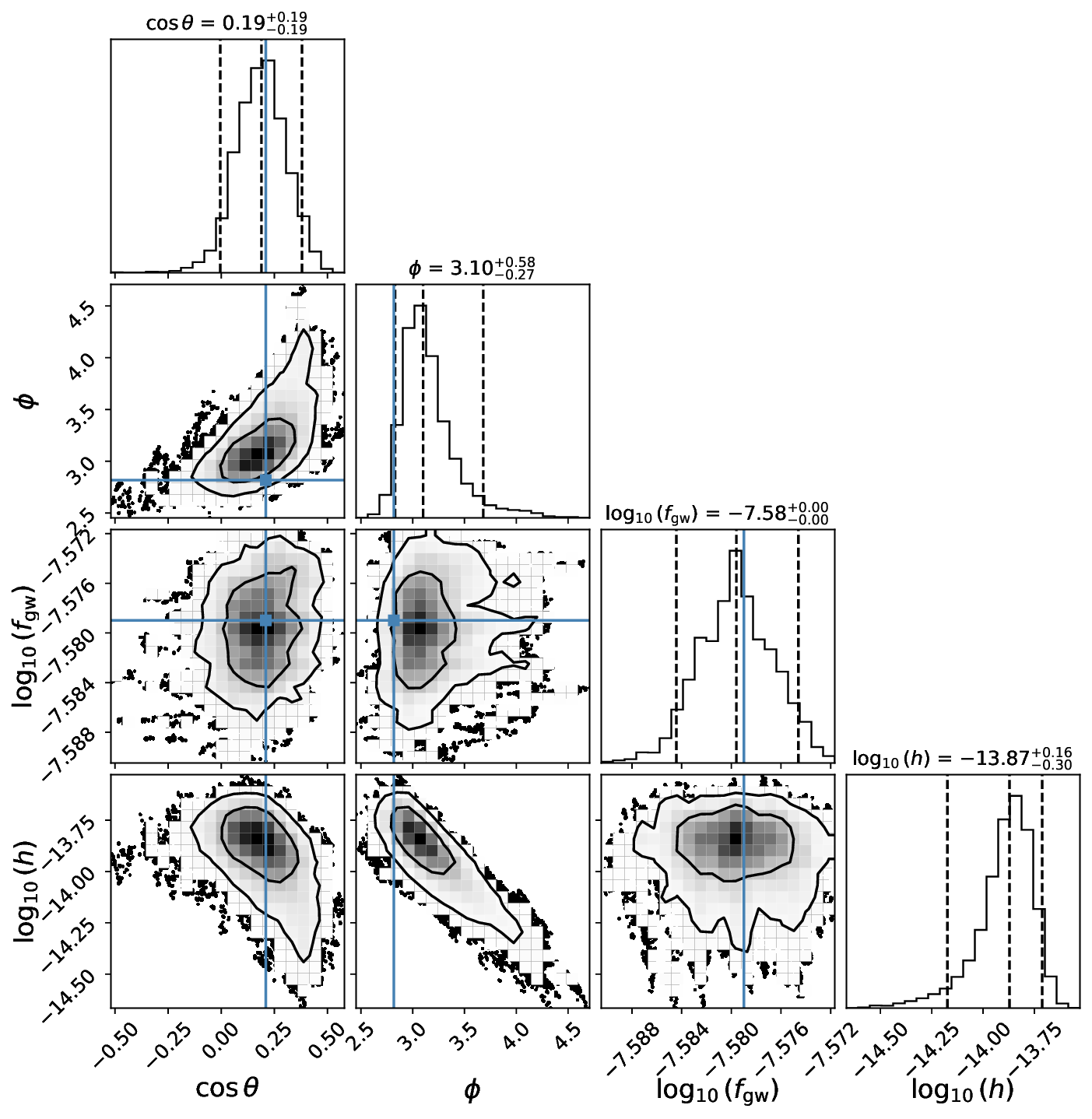}
  \caption{Same as Fig. \ref{fig:bayesian_avg} for the P-60 pulsar selection scheme.}
  \label{fig:bayesian_hf60U}
\end{figure}

\begin{figure}[htbp]
  \centering
  \includegraphics[width=0.48\textwidth,height=0.45\textheight,keepaspectratio]{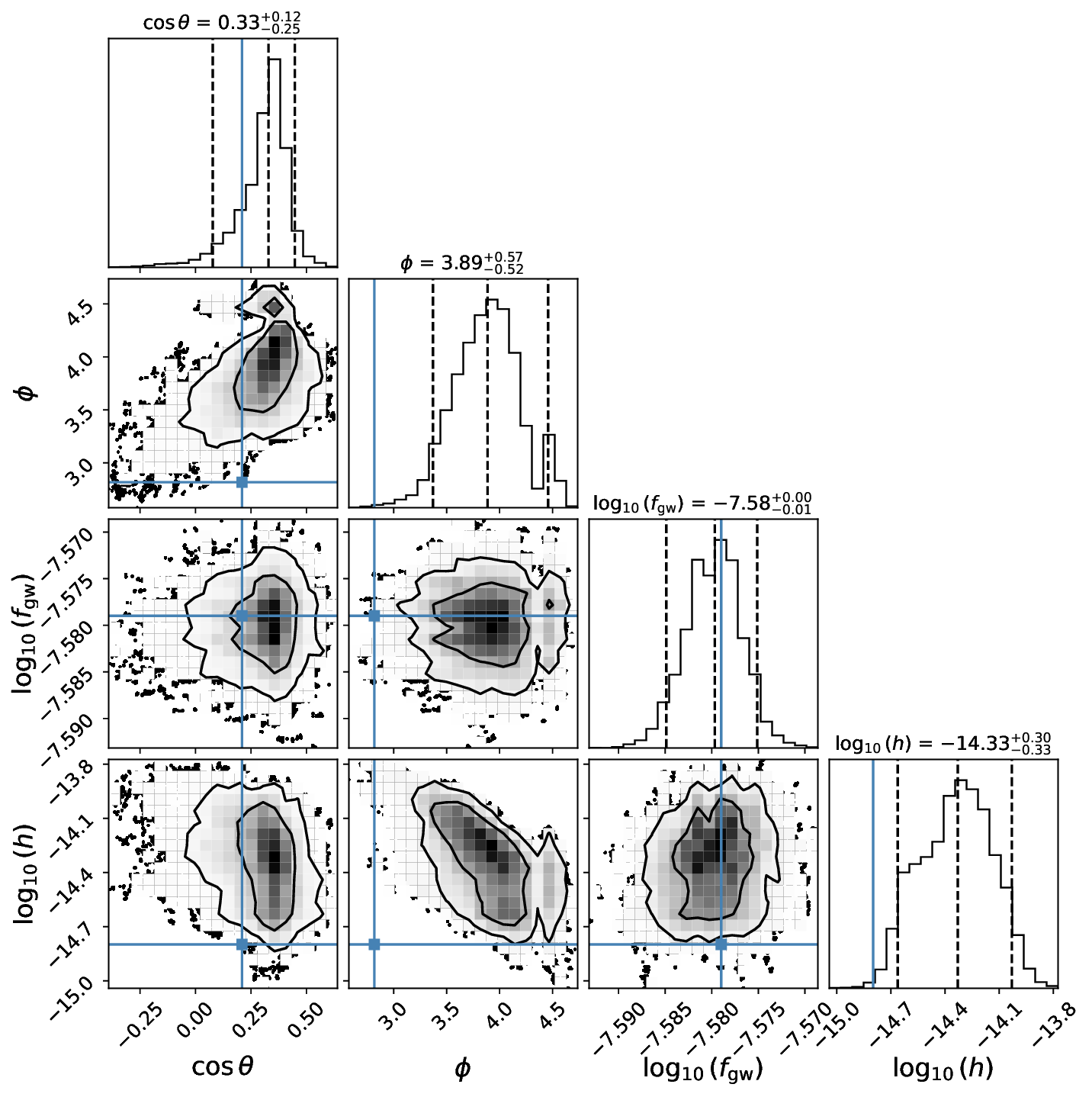}
  \caption{Same as Fig. \ref{fig:bayesian_avg} for the C-SNR-90 pulsar selection scheme.}
  \label{fig:bayesian_selected_90}
\end{figure}

\begin{figure*}[p]
  \centering
  \subfigure[C-ASNR-90]{
    \includegraphics[width=0.55\textwidth,height=0.4\textheight,keepaspectratio]{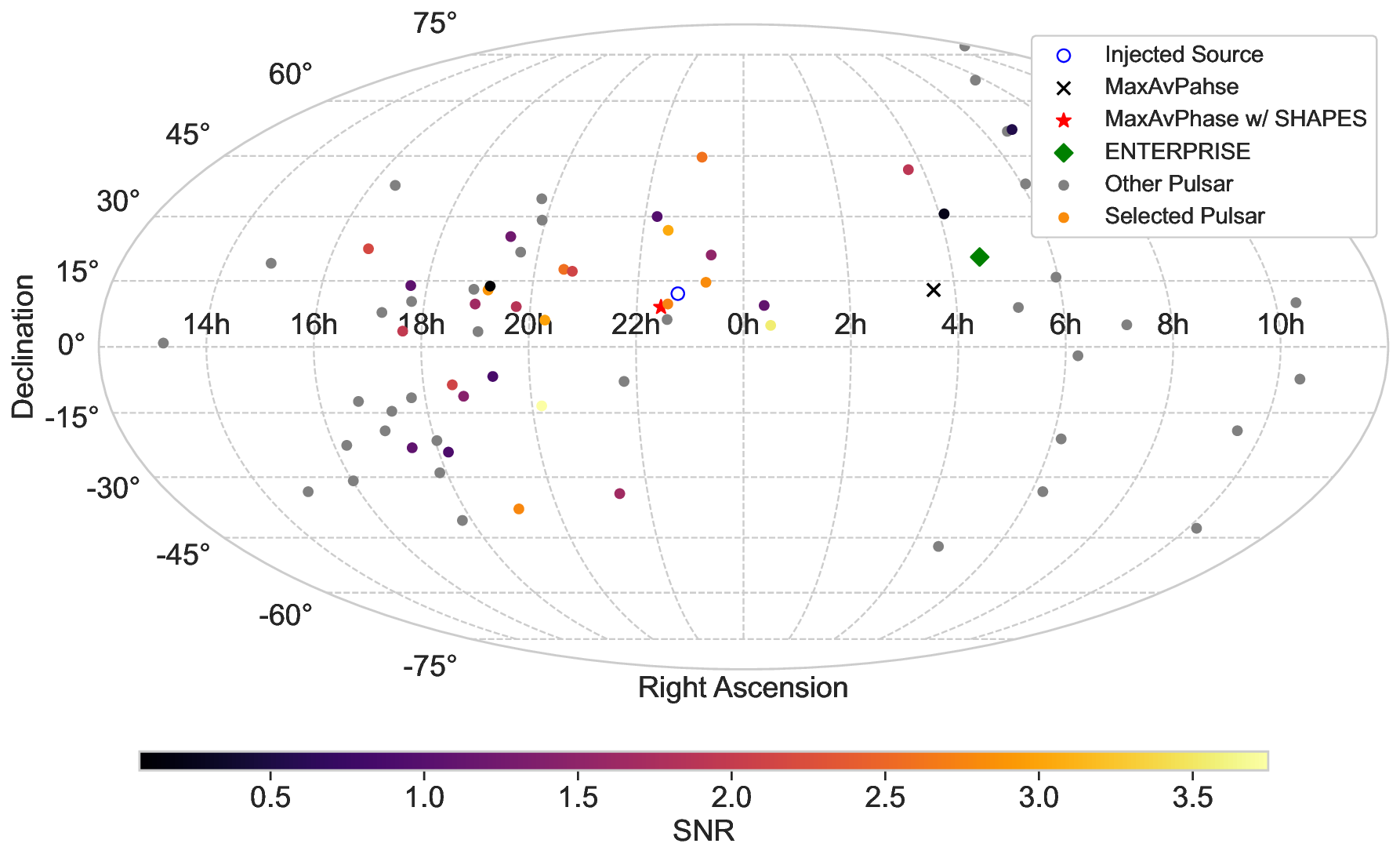}
    \label{fig:maxavphase_avg}
  }
  \vfill
  \subfigure[P-60]{
    \includegraphics[width=0.55\textwidth,height=0.4\textheight,keepaspectratio]{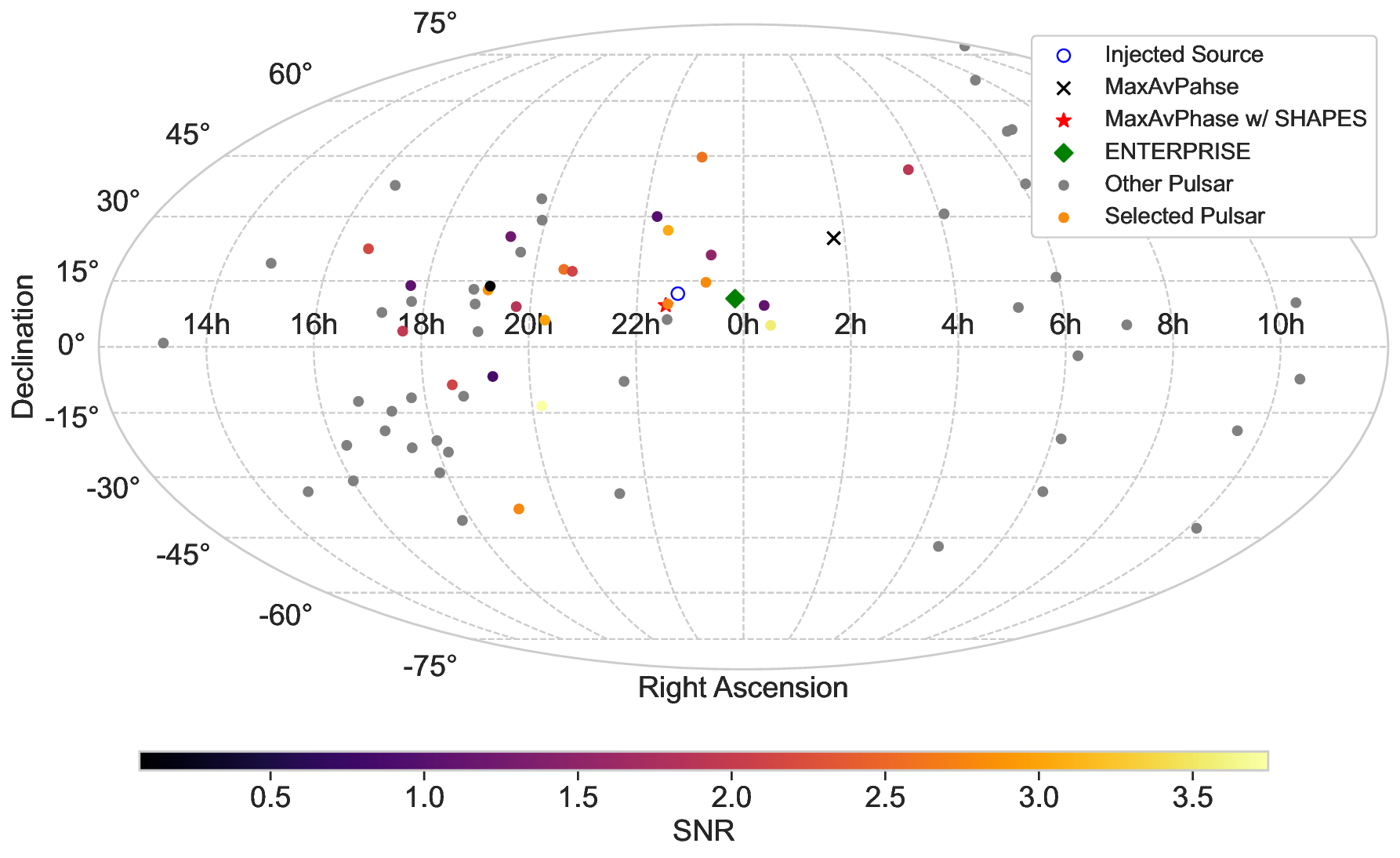}
    \label{fig:maxavphase_hf60u}
  }
  \vfill
  \subfigure[C-SNR-90]{
    \includegraphics[width=0.55\textwidth,height=0.4\textheight,keepaspectratio]{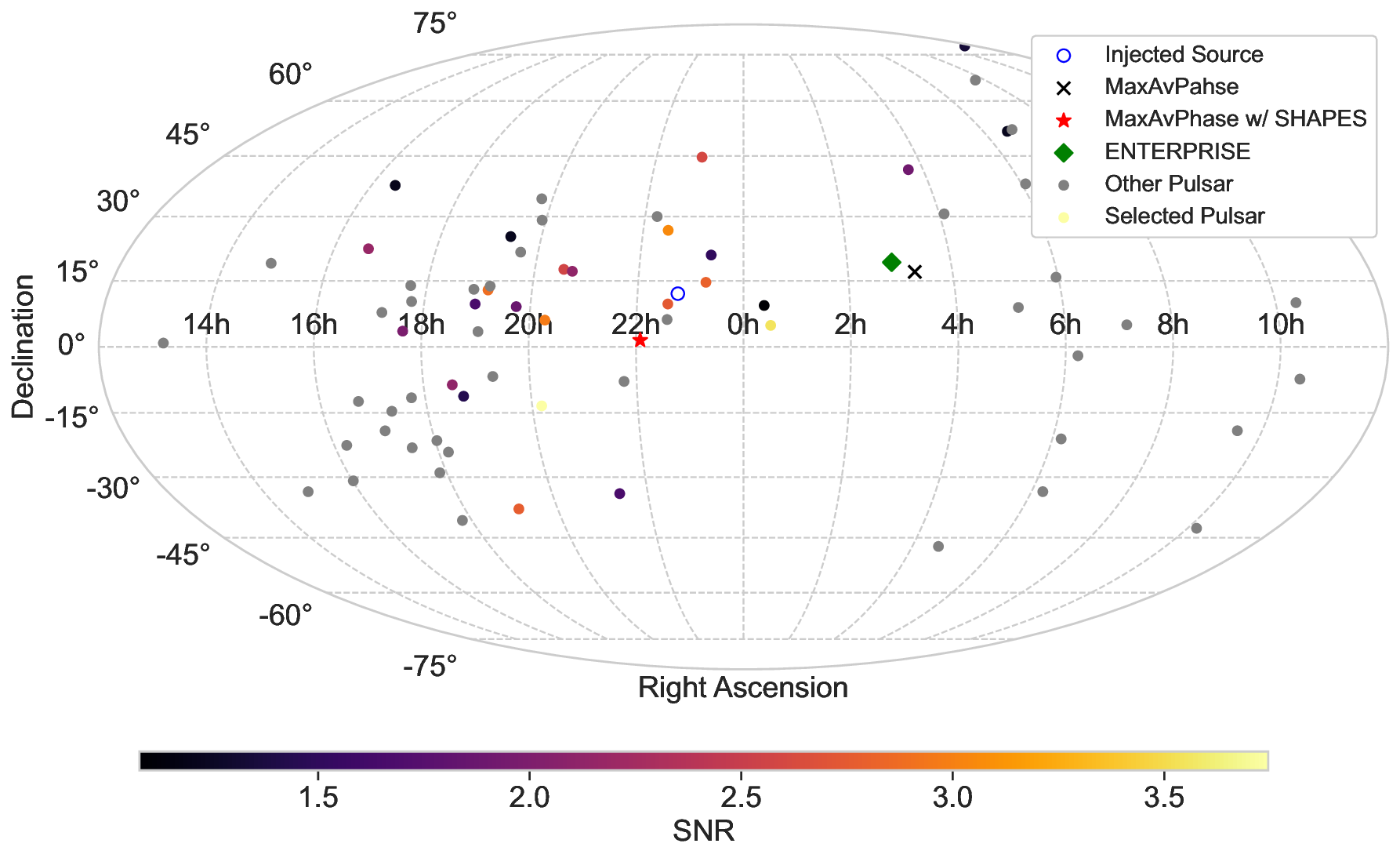}
    \label{fig:maxavphase_90}
  }
  \caption{The Bayesian and SM search results for different pulsar selection schemes. The Mollweide plots show the localization results for the CGW parameters for the following pulsar subsets: (a) C-ASNR-90, (b) P-60, and (c) C-SNR-90. Gray dots represent the pulsars that are not selected, while colored dots represent the selected pulsars, with the color indicating their SNR. The blue circle represents the injected CGW source, black cross represents the MaxAvPhase result, the red star represents the SM result, and the green diamond represents the Bayesian result.}
  \label{fig:bayesian_and_ms_search_results_comparison}
\end{figure*}

Fig.~\ref{fig:bayesian_and_ms_search_results_comparison} shows the localization results for the Bayesian and SM methods for the three optimized pulsar subsets (the full PTA result, where both methods failed to localize the source, is shown in Appendix~\ref{app:full_pta_results}). In the C-ASNR-90 and C-SNR-90 sets, the SM method has a better localization ability than the Bayesian method. When using the P-60 pulsars, both methods have comparable localization ability, for this data realization. It is worth noting that the Bayesian method performs similarly to the MaxAvPhase-only method for the C-ASNR-90 and C-SNR-90 sets, but performs better with the P-60 set. This is likely because the C-ASNR-90 and C-SNR-90 sets contain one more pulsar (B1855+09) with quite strong red noise than the P-60 set, which might imply that the CGW signal is entangled with the red noise for certain pulsars and the Bayesian model does not distinguish them, while using the SM method, the red noise is suppressed and the CGW signal is better localized.

\begin{figure*}[htbp]
  \centering
  \subfigure[CGW source location A]{
    \includegraphics[width=0.9\textwidth]{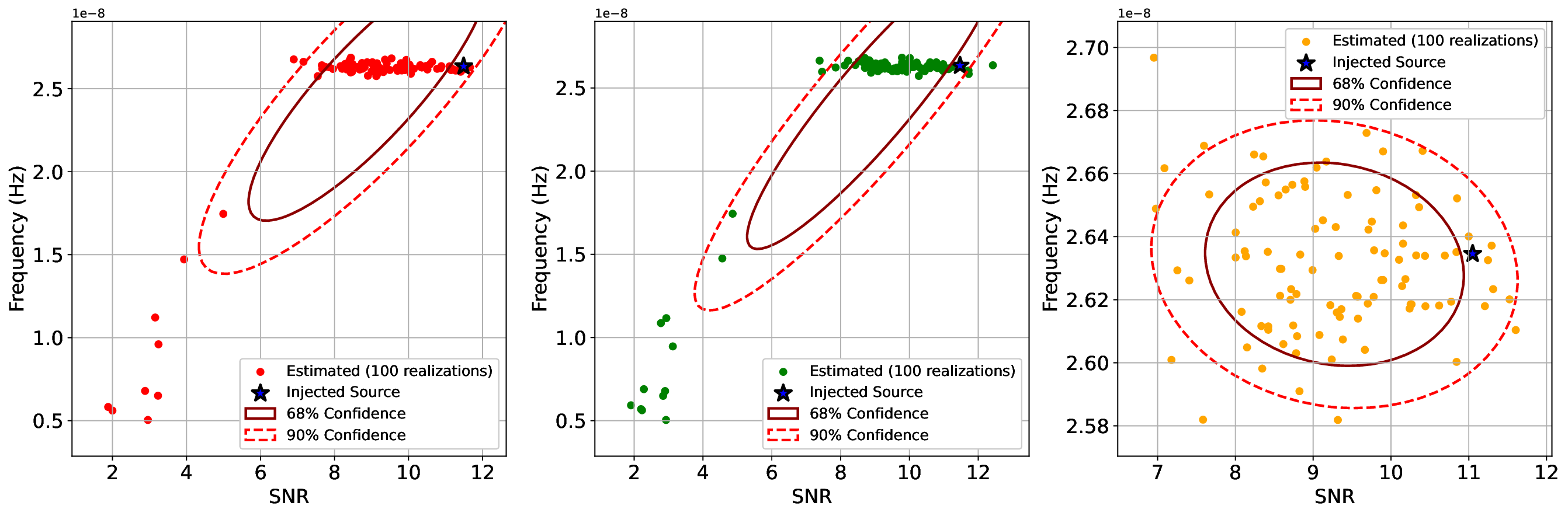}
    \label{fig:snr_frequency_analysis_separate_a}
  }
  \vfill
  \subfigure[CGW source location B]{
    \includegraphics[width=0.9\textwidth]{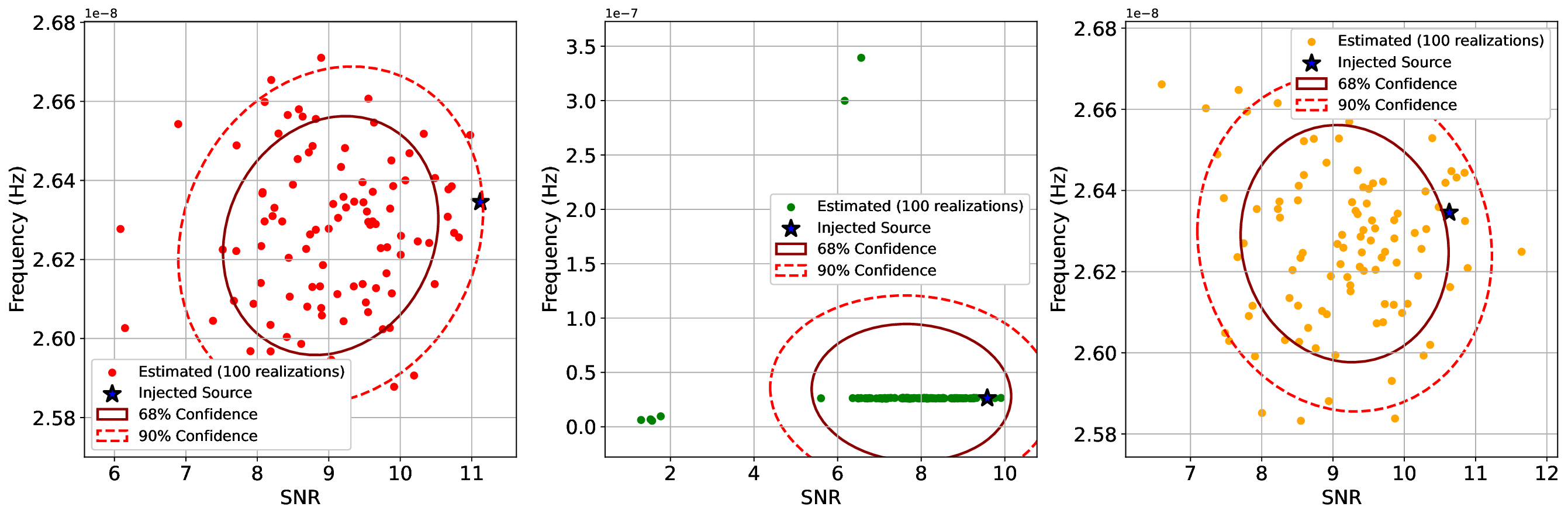}
    \label{fig:snr_frequency_analysis_separate_b}
  }
  \caption{SNR and frequency estimation performance for different pulsar selection schemes of the SM method, evaluated using 100 independent noise realizations. Panels (a) and (b) present the recovered SNR and frequency for the CGW source in location A and B, respectively. The left, middle, and right columns correspond to the C-ASNR-90, C-SNR-90, and P-60 schemes, respectively. Solid dots indicate the estimates obtained from the 100 realizations, and the star marks the injected values. Solid and dashed lines denote the 68\% and 90\% confidence contours, respectively.}
  \label{fig:snr_frequency_analysis_separate}
\end{figure*}

\subsection{Statistical study of performance \label{robustness_check_for_3_sets_of_optimized_pulsars}}
For a rigorous statistical study of the performance of the MS method, we use 100 noise realizations for each pulsar selection scheme. As discussed in Sec.~\ref{pulsar_optimization}, the subset of pulsars obtained from the selection schemes depends on the sky location and, for C-SNR-90, also on the frequency of the GW source. Therefore, we also consider an injected CGW signal at a different sky location B ((RA, DEC) = (5.71, -0.04)), while keeping all other parameters the same.

The selected pulsars for three schemes are shown in Table~\ref{tab:pulsar_difference2}. By comparing with Table~\ref{tab:pulsar_difference}, we have several interesting findings: (1) The P-60 set demonstrates the highest degree of compositional stability among the three groups. While there was significant internal restructuring—characterized by the reassignment of several pulsars (e.g., J1923+2515, J2043+1711) from the ``Common" category to the P-60 specific list—the overall integrity of the set remains largely intact. The P-60 set successfully retained these reclassified pulsars and incorporated the newly added pulsar J1455-3330, suggesting that the P-60 metric provides a robust baseline that is less sensitive to the fluctuations that affected the other criteria; (2) The full C-ASNR-90 set functions as a broad inclusion filter in the new distribution. Although it excluded several specific sources present in the original distribution (e.g., B1855+09, J0406+3039), it notably captured all pulsars that were removed from the ``Common" set (such as J0340+4130 and J2302+4442). This indicates that while the C-ASNR-90 metric filters out certain lower-quality outliers, it remains lenient enough to encompass pulsars that no longer meet the stricter convergence requirements of the Common or C-SNR-90 sets; (3) The C-SNR-90 set exhibits the most drastic reduction, indicative of a stringent tightening of the selection criteria or high sensitivity to the updated noise models. Unlike the other groups, C-SNR-90 failed to retain the pulsars downgraded from the ``Common" set (e.g., J0340+4130, J1923+2515) and simultaneously discarded the majority of its originally unique members (e.g., J0636+5128, J1125+7819). Consequently, the new C-SNR-90 set represents a highly refined, exclusive subset, retaining only the pulsars with the highest signal-to-noise reliability in the updated analysis.

\begin{figure*}[htbp]
  \centering
  \subfigure[CGW source in location A]{
    \includegraphics[width=0.8\textwidth]{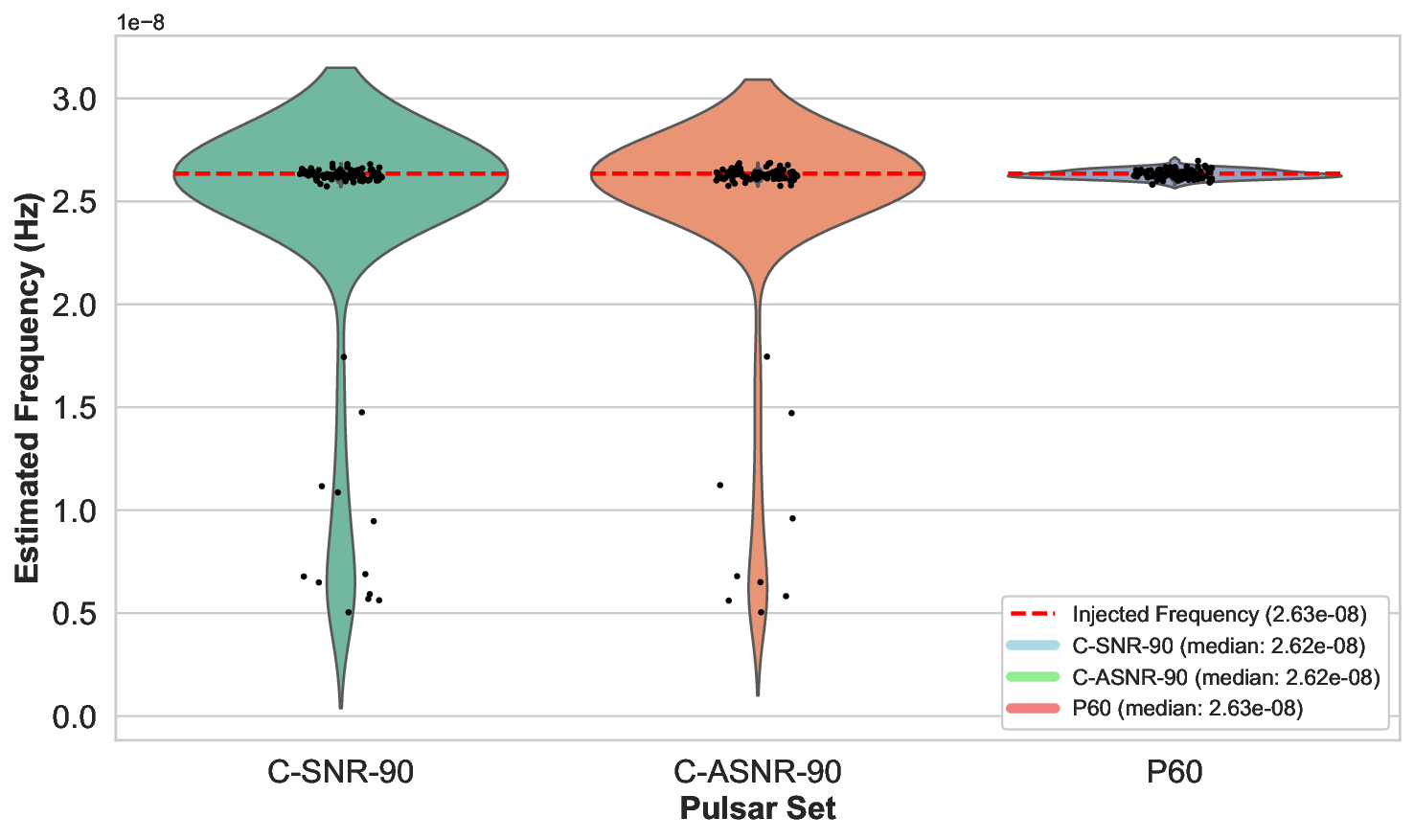}
  }
  \\
  \subfigure[CGW source in location B]{
    \includegraphics[width=0.8\textwidth]{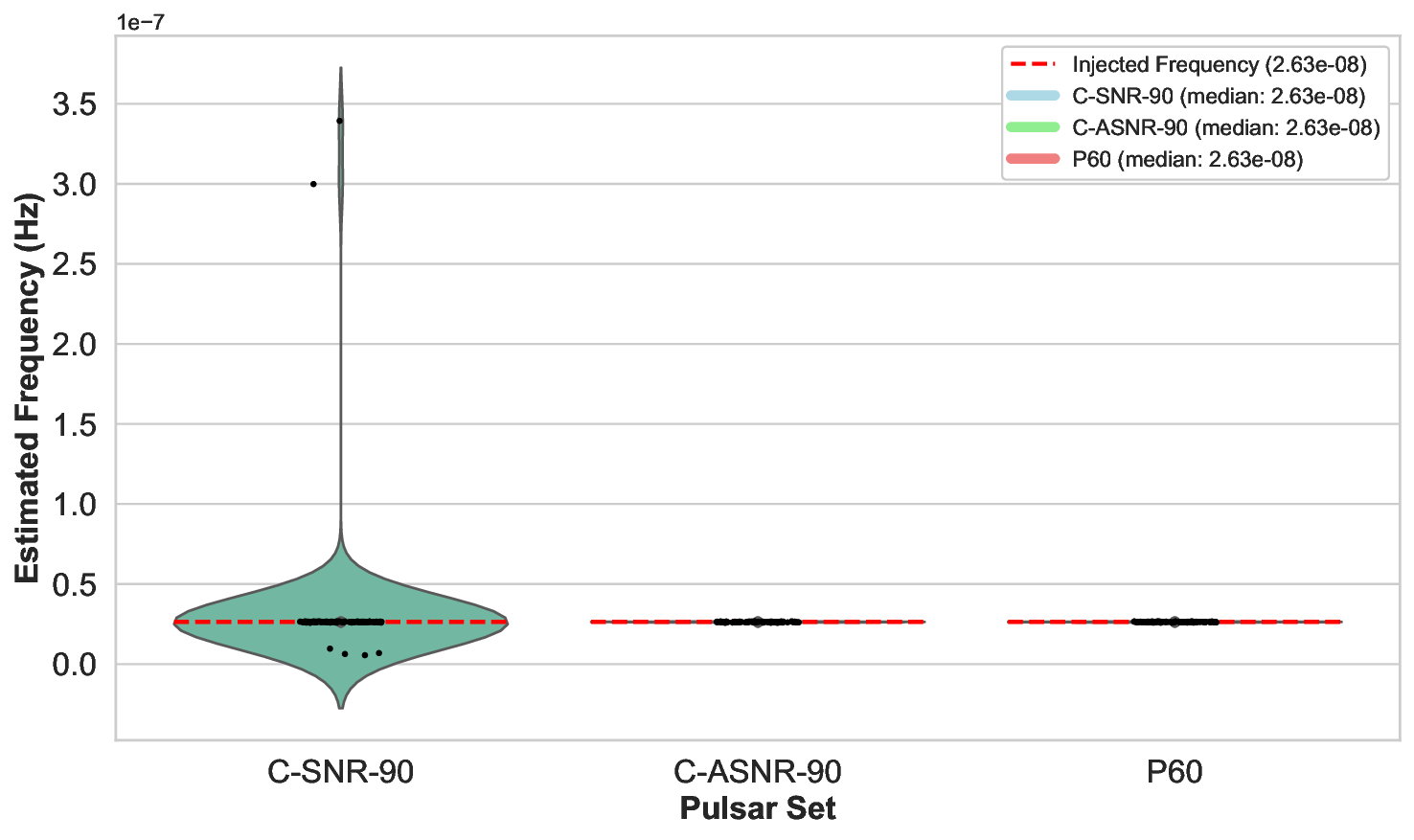}
  }
  \caption{Estimated frequency distributions across different pulsar selection schemes using the SM method for 100 noise realizations. Panel (a) and (b) show the estimated frequency distributions for the CGW source in location A and B, respectively. The red dashed line represents the injected frequency. The black dots are the estimated values of frequency and the shape of the violin represents the distribution density of the estimates.}
  \label{fig:frequency_error_analysis}
\end{figure*}

Next, we discuss the performance details of these three optimized pulsar sets. Fig.~\ref{fig:snr_frequency_analysis_separate} displays the estimated SNR-frequency plot for various pulsar selection schemes across 100 noise realizations. We observe that C-ASNR-90 performs better in location B, exhibiting only a few poorly estimated outliers. This is likely due to the exclusion of pulsars with strong red noise, such as B1855+09. In contrast, A-SNR-90 performs consistently poorly across two different sky locations, even though its selected pulsars vary significantly. Among the three sets, P-60 performs the best. It maintains stability in its selected pulsars and parameter estimation performance, regardless of the sky location. This is more clearly seen in Fig.~\ref{fig:frequency_error_analysis}, which illustrates the distribution of estimated frequencies across the three different pulsar sets. The violin plots for both C-SNR-90 and C-ASNR-90 exhibit similar morphological characteristics in location A, characterized by a broad probability distribution and a long tail extending towards lower frequencies. While the performance of C-ASNR-90 improves dramatically in location B with no outliers at all and reaches the same performance of P-60, C-SNR-90 still has several badly estimated outliers. The P-60 set displays a more concentrated distribution with negligible dispersion in location A and B. Its data points cluster tightly around the injected value, demonstrating superior precision and stability in the parameter estimation compared to the other configurations. It is worth to note that even though the A-CSNR-90 and C-SNR-90 sets may have outliers in the parameter estimation, their median values are still close to the injected values.

\begin{table}[h!]
  \centering
  \caption{As in Table~\ref{tab:pulsar_difference}, the pulsar subsets obtained using the three selection schemes described in Sec.~\ref{pulsar_optimization} for the GW source location B. }
  \label{tab:pulsar_difference2}
  \begin{tabular}{|l|l|l|}
    \hline
    \multicolumn{3}{|c|}{Common PSRs} \\
    \hline
    \multicolumn{3}{|c|}{J0023+0923} \\
    \multicolumn{3}{|c|}{J0030+0451*} \\
    \multicolumn{3}{|c|}{J1455-3330} \\
    \multicolumn{3}{|c|}{J1640+2224} \\
    \multicolumn{3}{|c|}{J1738+0333} \\
    \multicolumn{3}{|c|}{J1832-0836} \\
    \multicolumn{3}{|c|}{J1909-3744*} \\
    \multicolumn{3}{|c|}{J1910+1256} \\
    \multicolumn{3}{|c|}{J1918-0642} \\
    \multicolumn{3}{|c|}{J1944+0907} \\
    \multicolumn{3}{|c|}{J2010-1323} \\
    \multicolumn{3}{|c|}{J2017+0603} \\
    \multicolumn{3}{|c|}{J2033+1734} \\
    \multicolumn{3}{|c|}{J2214+3000} \\
    \multicolumn{3}{|c|}{J2229+2643} \\
    \multicolumn{3}{|c|}{J2234+0944} \\
    \multicolumn{3}{|c|}{J2317+1439} \\
    \multicolumn{3}{|c|}{J2322+2057} \\
    \hline
    \multicolumn{3}{|c|}{Different PSRs} \\
    \hline
    J1751-2857 & J1741+1351 & J0340+4130 \\
    J1811-2405 & J1911+1347 & J1741+1351 \\
    J2124-3358 & J1923+2515 & J1811-2405 \\
    & J2043+1711 & J1843-1113 \\
    & J2302+4442 & J1911+1347 \\
    & & J1923+2515 \\
    & & J2043+1711 \\
    & & J2124-3358 \\
    & & J2302+4442 \\
    \hline
    \multicolumn{1}{|c|}{\textbf{C-SNR-90}} & \multicolumn{1}{c|}{\textbf{P-60}} & \multicolumn{1}{c|}{\textbf{C-ASNR-90}} \\
    \hline
  \end{tabular}
\end{table}

\begin{figure*}[htbp]
  \centering
  \subfigure[CGW source in location A]{
    \includegraphics[width=0.8\textwidth]{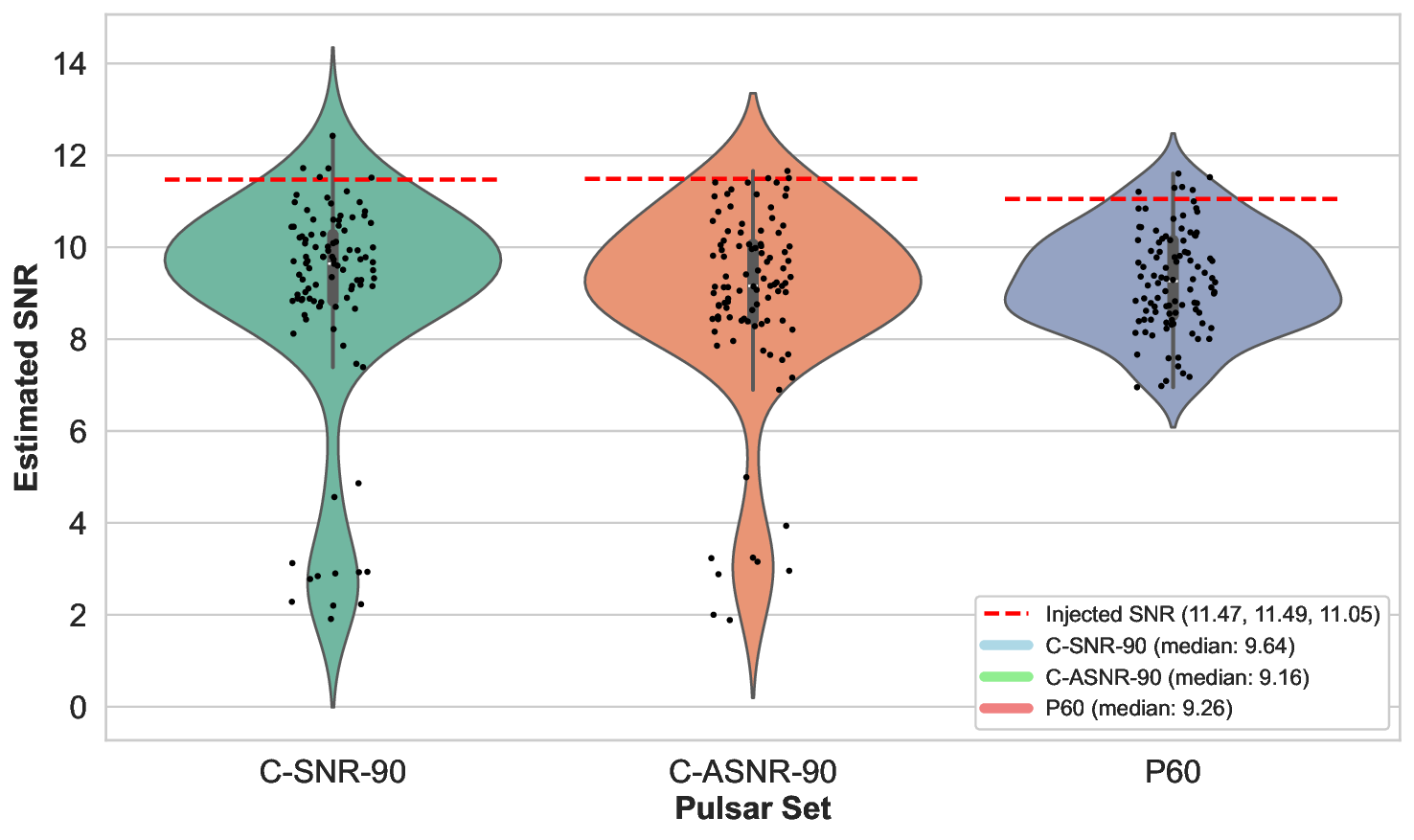}
  }
  \\
  \subfigure[CGW source in location B]{
    \includegraphics[width=0.8\textwidth]{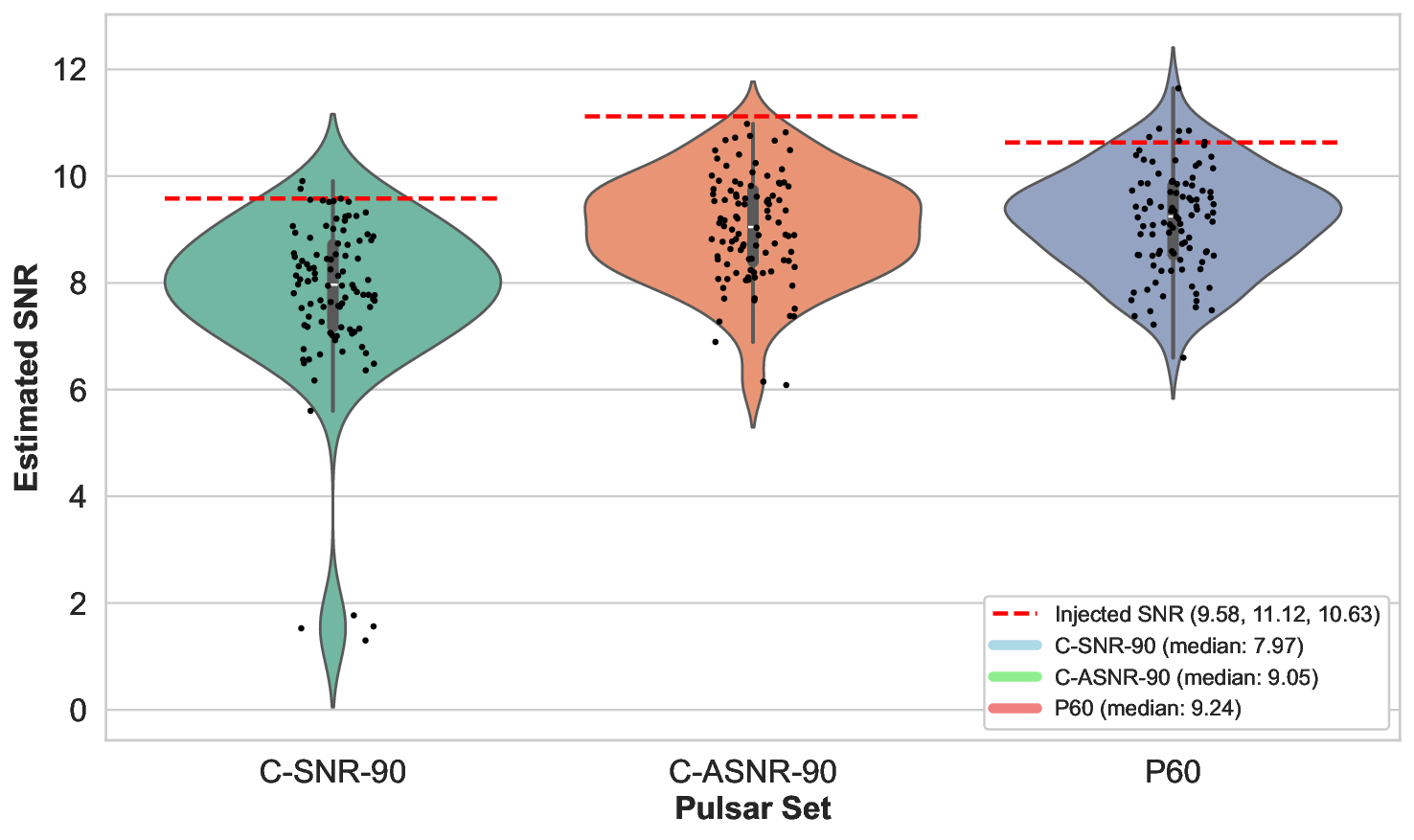}
  }
  \caption{Estimated SNR distributions across different pulsar selection schemes using the SM method for 100 noise realizations. Panel (a) and (b) show the estimated SNR distributions for the CGW source in location A and B, respectively. The red dashed line represents the injected SNR calculated using the corresponding pulsar subsets. The black dots are the estimated values of SNR and the shape of the violin represents the distribution density of the estimates. }
  \label{fig:snr_violin_comparison}
\end{figure*}

Consistent with the performance trends observed in frequency estimation, the SNR analysis presented in Fig.~\ref{fig:snr_violin_comparison} reveals a similar hierarchy among the pulsar sets, albeit with less pronounced visual distinctions. The C-SNR-90 and C-ASNR-90 sets continue to exhibit a broader dispersion and a more significant deviation from the injected values in location A. While the P-60 set still demonstrates relatively superior stability, its distribution is slightly more condensed around its median. However, a common limitation affects all three configurations: unlike the precise frequency recovery seen in the P-60 set, the SNR estimates across the board suffer from a systematic downward bias, with all recovered medians (approx. 9.2--9.6) falling consistently short of the injected values (approx. 11.0--11.5).

\begin{figure*}[htbp]
  \centering
  \includegraphics[width=\textwidth]{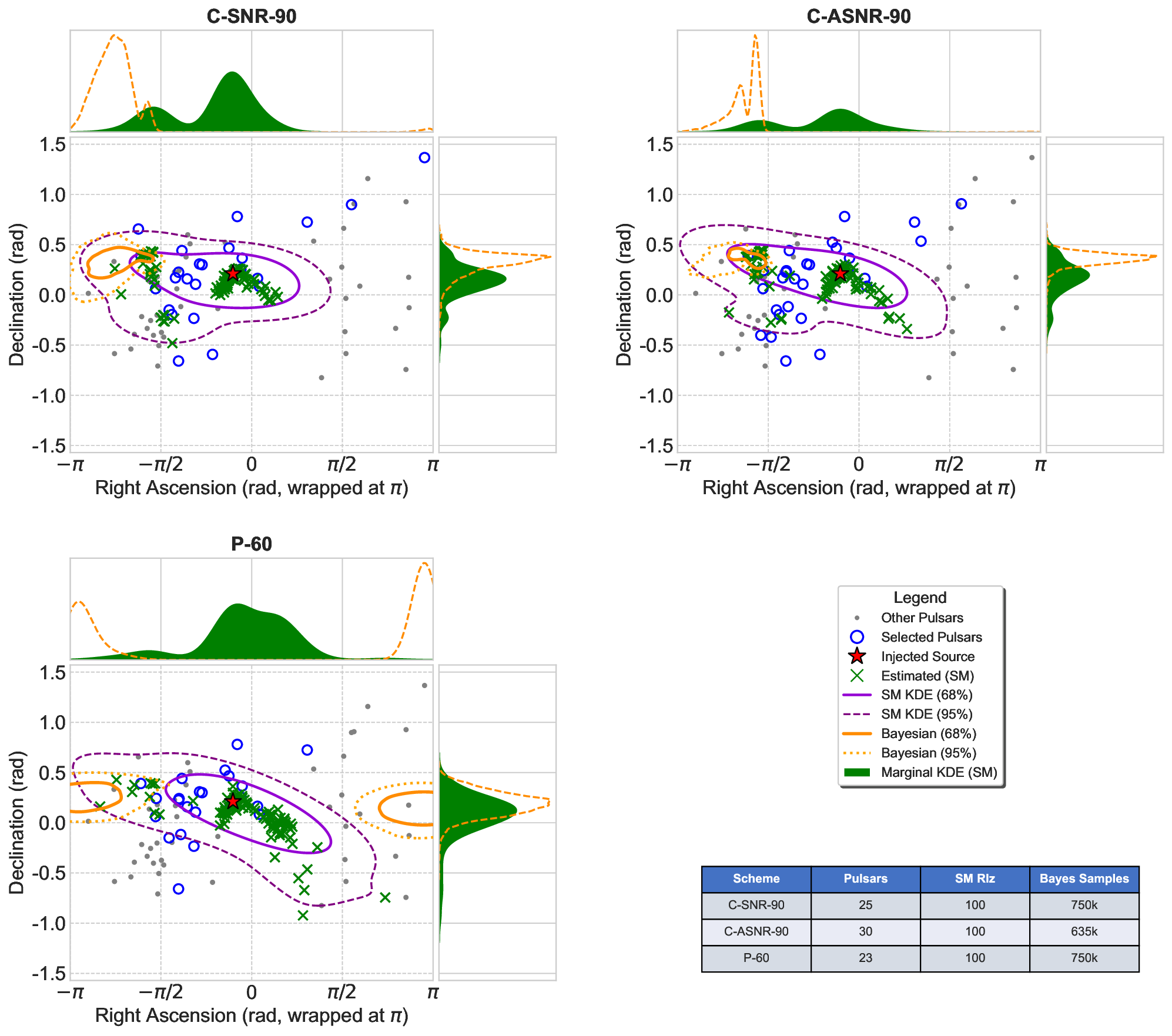}
  \caption{Sky localization comparison for the GW source in location A across different pulsar selection schemes using the SM and Bayesian methods. The upper left, upper right, and lower left panels display the results for the C-SNR-90, C-ASNR-90, and P-60 optimized pulsar sets, respectively. The lower right panel provides a legend and summary of all three schemes. Marginal plots at the top and right of each panel show the KDE of the SM estimates for the sky locations. Gray dots indicate unselected pulsars, while blue circles denote pulsars selected by the corresponding scheme. The red star marks the injected GW source, and green crosses represent the SM estimates from 100 noise realizations. Solid and dashed lines indicate the 68\% and 95\% confidence regions, respectively, with different colors corresponding to different methods. The Bayesian contours are derived from a single data realization, with the samples shown after burn-in from a total of $10^6$ samples.}
  \label{fig:sky_localization_comparison}
\end{figure*}

\begin{figure*}[htbp]
  \centering
  \includegraphics[width=\textwidth]{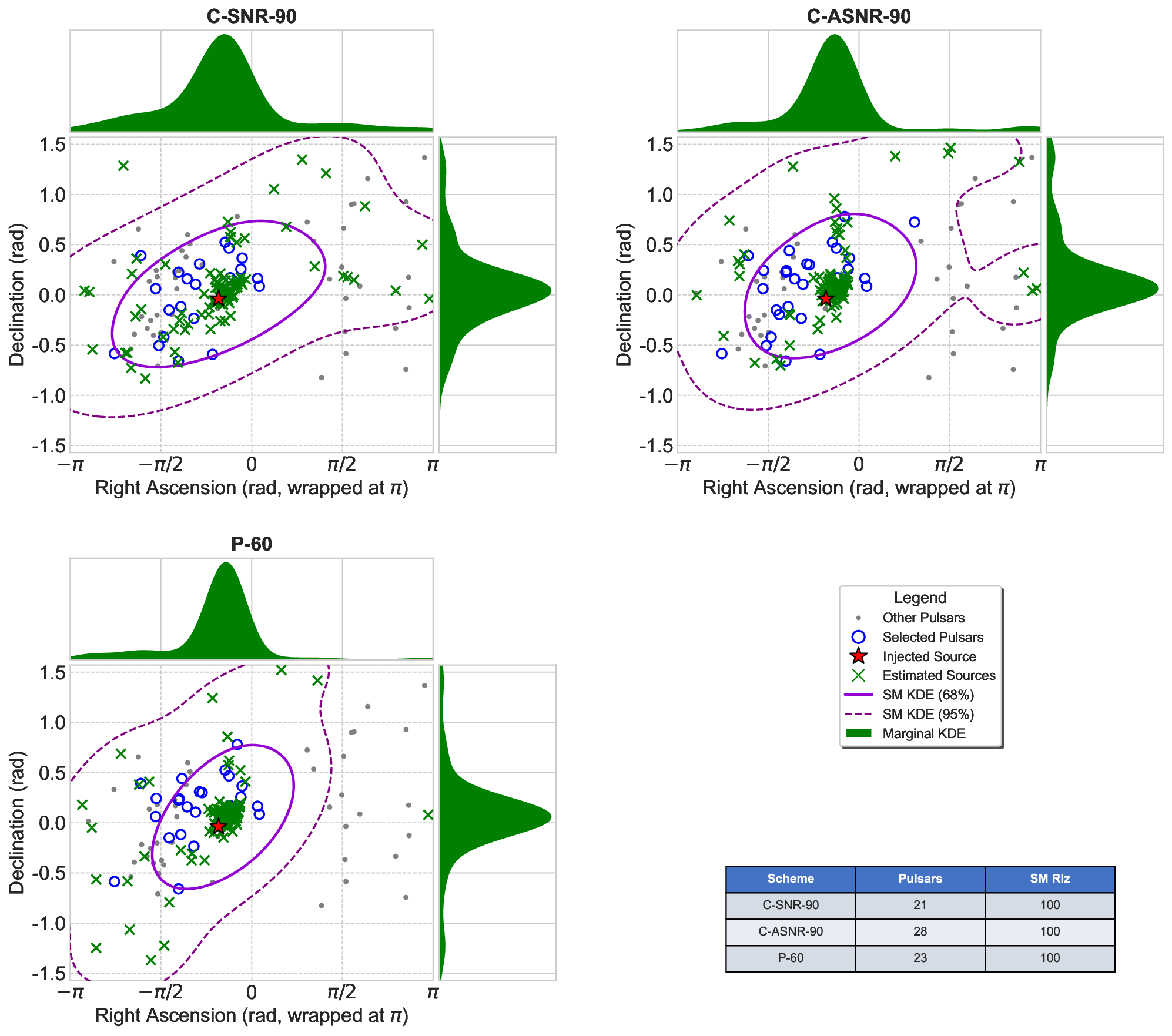}
  \caption{As in Fig.~\ref{fig:sky_localization_comparison}, the sky localization comparison for GW source in location B across different pulsar selection schemes using the SM and Bayesian methods.}
  \label{fig:sky_localization_comparison_2}
\end{figure*}

Fig.~\ref{fig:sky_localization_comparison} and Fig.~\ref{fig:sky_localization_comparison_2} show the sky localization results for the three sets of pulsars in the two different sky locations. The marginal plots at the top and right show the kernel density estimation (KDE)~\cite{rosenblattRemarksNonparametricEstimates1956,parzenEstimationProbabilityDensity1962} of the sky locations for the SM method. Due to computational limitations, we only show the Bayesian contours for one sky location for comparison. The figures indicate that the Bayesian method provides a more stringent constraint on the sky location than the SM method, but it exhibits a slight bias in the best sky location estimate, this may be due to the ET-only search~\cite{zhuDetectionLocalizationContinuous2016a}. A recent study~\cite{grunthalRoleDistantPulsars2025a} shows that including distant pulsars may improve the performance of the ET-only search using the Bayesian method. Although the SM method offers better localization of the true sky locations, it provides a looser constraint on them. Furthermore, by comparing the results for the two sky locations, we find that in location A, the C-ASNR-90 and C-SNR-90 seem to have a bi-modal distribution, which is consistent with the Bayesian results. This may be due to the effects of pulsar B1855+09, which is included in the C-ASNR-90 and C-SNR-90 sets for location A, but not in the P-60 set. In location B, however, this pattern disappears, which is likely because the pulsar B1855+09 is not included in the C-ASNR-90 and C-SNR-90 sets for location B. Hence, we can draw the conclusion that pulsars with strong red noise significantly impact the ability to localize the GW sources for both the SM and the Bayesian methods.

\section{Conclusion \label{conclusion}}
In this paper, we have presented and validated a novel frequentist method for the detection of CGWs in PTA data. The central innovation of our work is the development of the SM method, which integrates red noise suppression using \texttt{SHAPES} with the selection of optimal pulsar subsets. Unlike conventional Bayesian techniques that necessitate complex, explicit modeling and time-consuming sampling of red noise for each pulsar, our SM method effectively suppresses intrinsic red noise through a non-parametric adaptive spline fitting process. This approach circumvents the intricate noise characterization stage, thereby providing a new, agile, and computationally efficient pathway for CGW searches.

To demonstrate the efficacy of our method, we performed analyses on a simulated dataset based on the NANOGrav 15-year data, into which a CGW signal with a signal-to-noise ratio of approximately 10 was injected. Our results show that the SM method, particularly when coupled with an optimized pulsar selection strategy, such as the P-60 scheme, robustly recovers the injected signal parameters. Under these optimal conditions, we achieved a relative SNR error of 7.84\% and a relative frequency error of only 0.072\%. These outcomes are highly comparable to those obtained from a full Bayesian analysis, confirming the accuracy and reliability of our method.

In order to address the computational challenges posed by large-scale arrays and the non-uniform distribution of sensitivity across the network, we evaluated three distinct pulsar optimization schemes: Cumulative SNR-90 (C-SNR-90), Cumulative Average SNR-90 (C-ASNR-90), and the Persistence-60 (P-60). While the C-SNR-90 and C-ASNR-90 schemes effectively identified high-contribution pulsars; however, our comparative analysis revealed their susceptibility to strong intrinsic red noise, which occasionally compromised sky localization accuracy and led to broader parameter dispersion. This highlights a fundamental limitation encountered by both Bayesian and frequentist frameworks: when the bandwidth of the intrinsic red noise overlaps significantly with the signal frequency and its amplitude substantially exceeds that of the signal, effective information extraction becomes infeasible. Consequently, including such noisy pulsars in the full PTA analysis obscures the signal, leading to suboptimal search results. This underscores the necessity of employing improved pulsar selection strategies to filter out these pulsars, thereby ensuring that the analysis is driven by high-quality data. In contrast, the P-60 scheme demonstrated superior robustness by rigorously filtering out pulsars with unstable noise characteristics across multiple frequency realizations, thereby retaining a highly stable subset of the array. This “quality over quantity” approach not only minimized the computational burden, but also yielded the most precise parameter estimation. These results effectively match the detection fidelity of the full array while significantly enhancing the scalability of the search pipeline.

The substantial improvement in computational efficiency positions our method as a scalable tool, well-suited for the large-scale datasets expected from future PTA experiments. Currently, our method is particularly advantageous for targeted searches where the source location is constrained. Another limitation of the current implementation is that the \texttt{SHAPES} algorithm requires evenly-sampled data. Although the underlying mathematical framework of B-splines is fully compatible with irregular sampling, the current code is optimized for uniform grids. We plan to develop a generalized version of \texttt{SHAPES} capable of handling non-uniformly sampled data in future updates. However, its direct application to all-sky blind searches presents challenges, as the optimal pulsar subset depends on the source's sky location. To address this and extend the method's versatility, future work will explore advanced selection strategies, such as a sky-grid strategy, where optimal pulsar subsets are pre-determined for specific grid cells and dynamically switched during the continuous gravitational wave search. We also note recent works on targeted CGW searches~\cite{agarwalNANOGrav15Yr2025,tianTargetedSearchIndividual2025} using Bayesian methods. These studies provide more stringent limits on the 95\% upper limit of the CGW characteristic strain and chirp mass. It would be straightforward to implement our method for targeted CGW searches.

The data that support the findings of this article are openly available~\cite{qian_2025_18092637}.

\section{Acknowledgment \label{acknowledgment}}
Y.W. gratefully acknowledges support from the National Key Research and Development Program of China (No. 2022YFC2205201 and No. 2023YFC2206702) and Major Science and Technology Program of Xinjiang Uygur Autonomous Region (No. 2022A03013-4). 
We acknowledge the High Performance Computing Platform at Huazhong University of Science and Technology for providing computational resources.
The authors thank the anonymous referee for helpful comments and suggestions. 

\appendix

\section{Bayesian analysis setup and full PTA results}
\label{app:bayesian_setup}
\label{app:full_pta_results}

\subsection{Detailed Bayesian running setup}

We used the default \texttt{PTMCMCSampler} parameters for all pulsar sets, including the full PTA: $\mathrm{SCAMweight}=30$, $\mathrm{AMweight}=20$, $\mathrm{DEweight}=50$, and $\mathrm{T}_{\mathrm{max}}=1$. We ran the MCMC for $10^7$ steps with a thinning factor of 10, and using OpenMP with 8 threads.

Additionally, we enabled the pulsar term in the \texttt{ENTERPRISE} model, but it did not improve the results. This is likely because the additional parameters increased the model's dimensionality, making MCMC convergence more difficult.

During the MCMC runs for the full PTA, we encountered numerical singularities where the likelihood function occasionally spiked to extremely large values, causing the chains to stall. This issue is likely attributable to strong red noise in certain pulsars. To address this, we applied a clipping procedure to remove these outlier likelihood values before generating the posterior distributions. The resulting corner plot is presented in Fig.~{\ref{fig:bayesian_full_app}}.

\subsection{Full PTA analysis results}

When the full PTA (all 68 pulsars) is used in the analysis, both the SM and Bayesian methods exhibit significantly degraded performance compared to the optimized pulsar subsets discussed in Sec.~\ref{bayesian_and_sm_search_results}. Table~\ref{tab:full_pta_error} summarizes the parameter estimation errors for the full PTA configuration. The SM method yields relative errors of 40.27\% in SNR, 20.87\% in characteristic strain, and 24.52\% in frequency, while the Bayesian method produces a characteristic strain error of 9.582\% and a frequency error of 238.3\%.

\begin{table}[htbp]
  \centering
  \caption{Relative errors of parameter estimation for the full PTA configuration.}
  \label{tab:full_pta_error}
  \begin{tabular}{cccc}
    \toprule
    \textbf{Methods} & $\frac{\delta \mathrm{SNR}}{\mathrm{SNR}}$ & $\frac{\delta \log_{10} h}{\log_{10} h}$ & $\frac{\delta f}{f}$ \\
    \midrule
    MaxAvPhase & 615.4\% & 22.31\% & 80.70\% \\
    SM & 40.27\% & 20.87\% & 24.52\% \\
    Bayesian & -- & 9.582\% & 238.3\% \\
    \bottomrule
  \end{tabular}
\end{table}

This deterioration arises because pulsars with strong intrinsic red noise, such as J1643-1224, J1705-1903, and J1745-1017, retain significant residual power in the low-frequency band---often 2--3 orders of magnitude higher than the injected signal---even after \texttt{SHAPES} detrending (see Appendix~\ref{app:more_LSPs} for the detailed Lomb-Scargle periodograms). These noisy pulsars, despite contributing little theoretical SNR, effectively mask the true signal during the search. The fact that both frequentist and Bayesian methods struggle with the full array underscores a critical finding: the quality of pulsars included in the analysis is more important than their quantity for CGW searches using the PTA.

\begin{figure*}[htbp]
  \centering
  \includegraphics[width=0.9\textwidth,height=0.5\textheight,keepaspectratio]{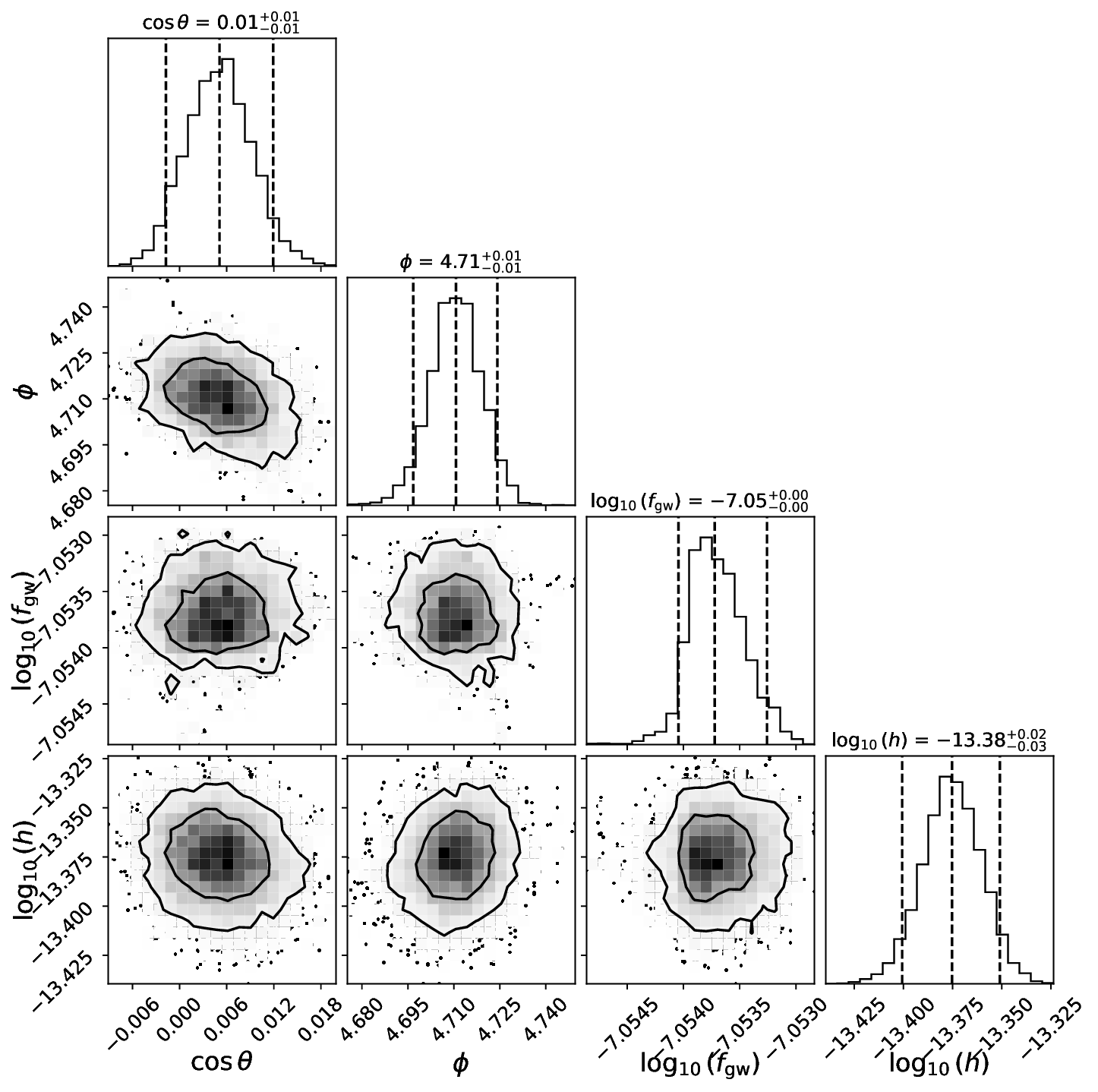}
  \caption{Bayesian search results for the full PTA for one data realization. The corner plot shows the 1D marginal histograms with 5\%, 50\%, and 95\% quantiles, and the 2D joint marginal distributions for the CGW parameters, including frequency, sky location, and characteristic strain. The outlier likelihood values caused by numerical singularities have been clipped before generating the posterior distributions.}
  \label{fig:bayesian_full_app}
\end{figure*}

\begin{figure*}[htbp]
  \centering
  \includegraphics[width=0.9\textwidth]{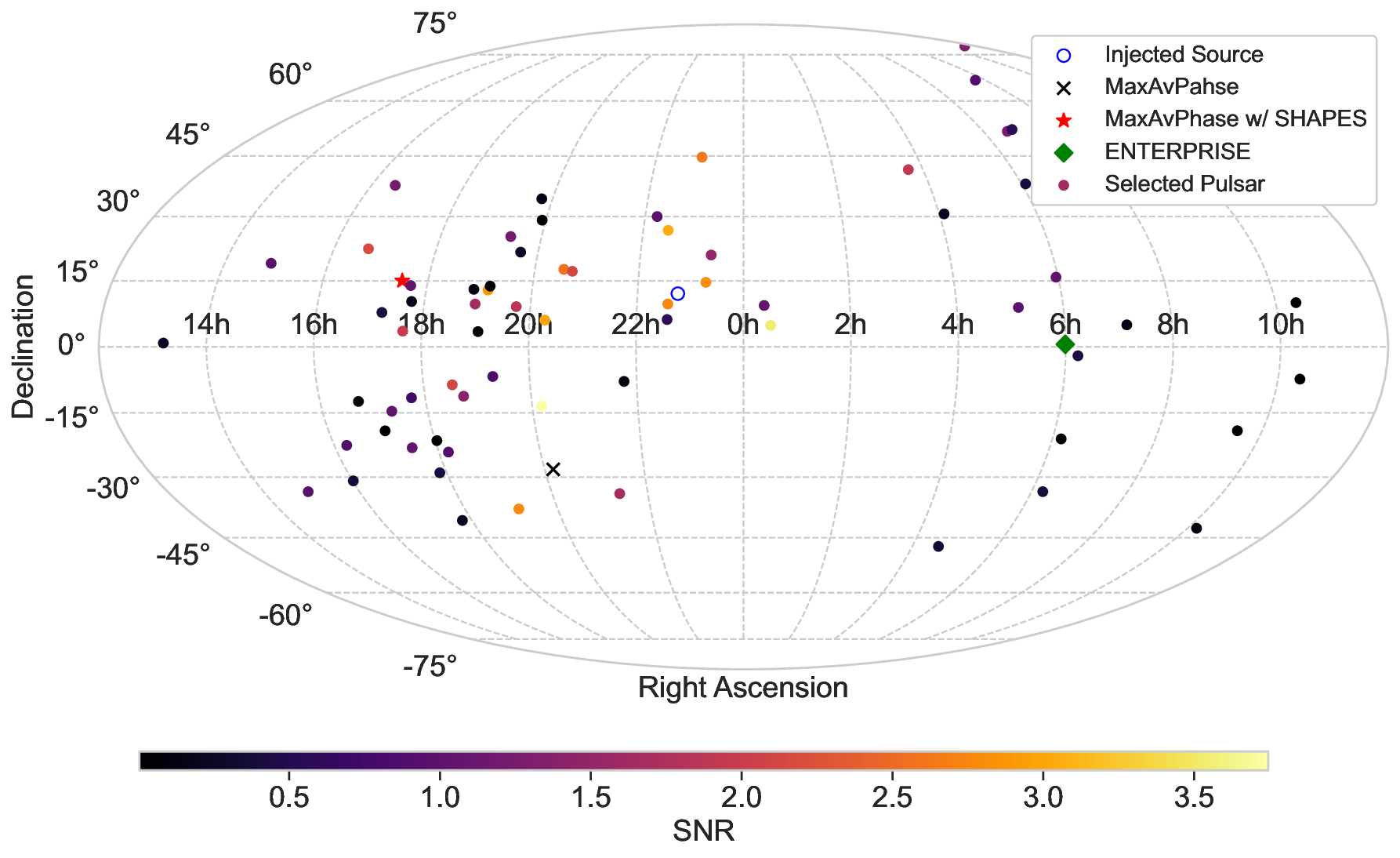}
  \caption{Sky localization result for the full PTA for one data realization. The Mollweide plot shows the localization results for the CGW parameters. Gray dots represent all pulsars, and colored dots indicate their SNR contributions. The blue circle represents the injected CGW source, the black cross represents the MaxAvPhase result, the red star represents the SM result, and the green diamond represents the Bayesian result.}
  \label{fig:maxavphase_full_app}
\end{figure*}

Fig.~\ref{fig:bayesian_full_app} shows the Bayesian corner plot for the full PTA, where the source parameters are poorly explored and the posterior distributions fail to recover the injected values. Fig.~\ref{fig:maxavphase_full_app} shows the corresponding sky localization result, confirming that neither the SM nor the Bayesian method can localize the CGW source when the full PTA is used.

\section{Additional Lomb-Scargle Periodograms}
\label{app:more_LSPs}

\begin{figure*}[htbp]
  \centering
  \subfigure[PSR J1643-1224]{
    \label{fig:app_J1643-1224_LSP}
    \includegraphics[width=0.45\textwidth]{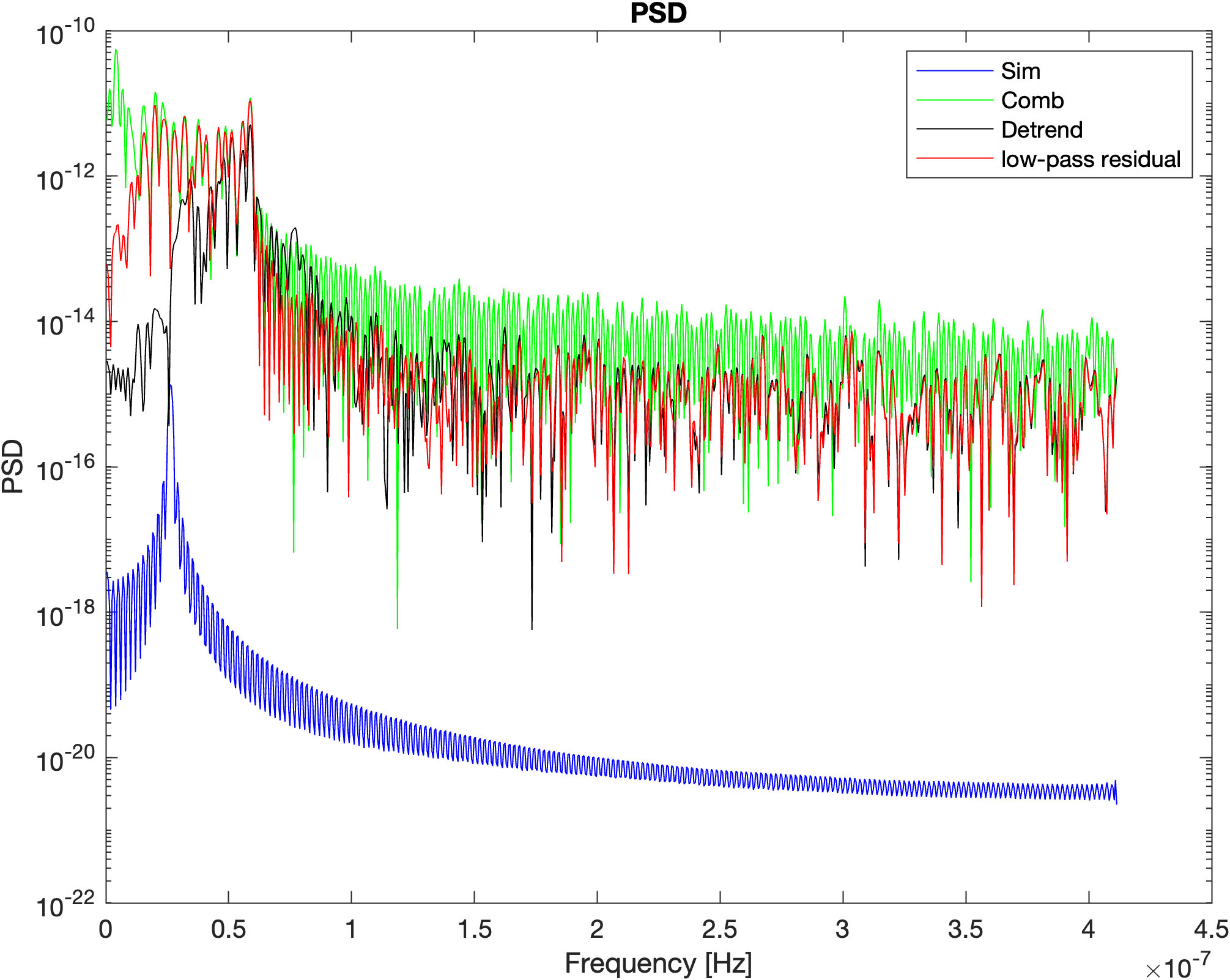}
  }
  \hfill
  \subfigure[PSR J1705-1903]{
    \label{fig:app_J705-1903_LSP}
    \includegraphics[width=0.45\textwidth]{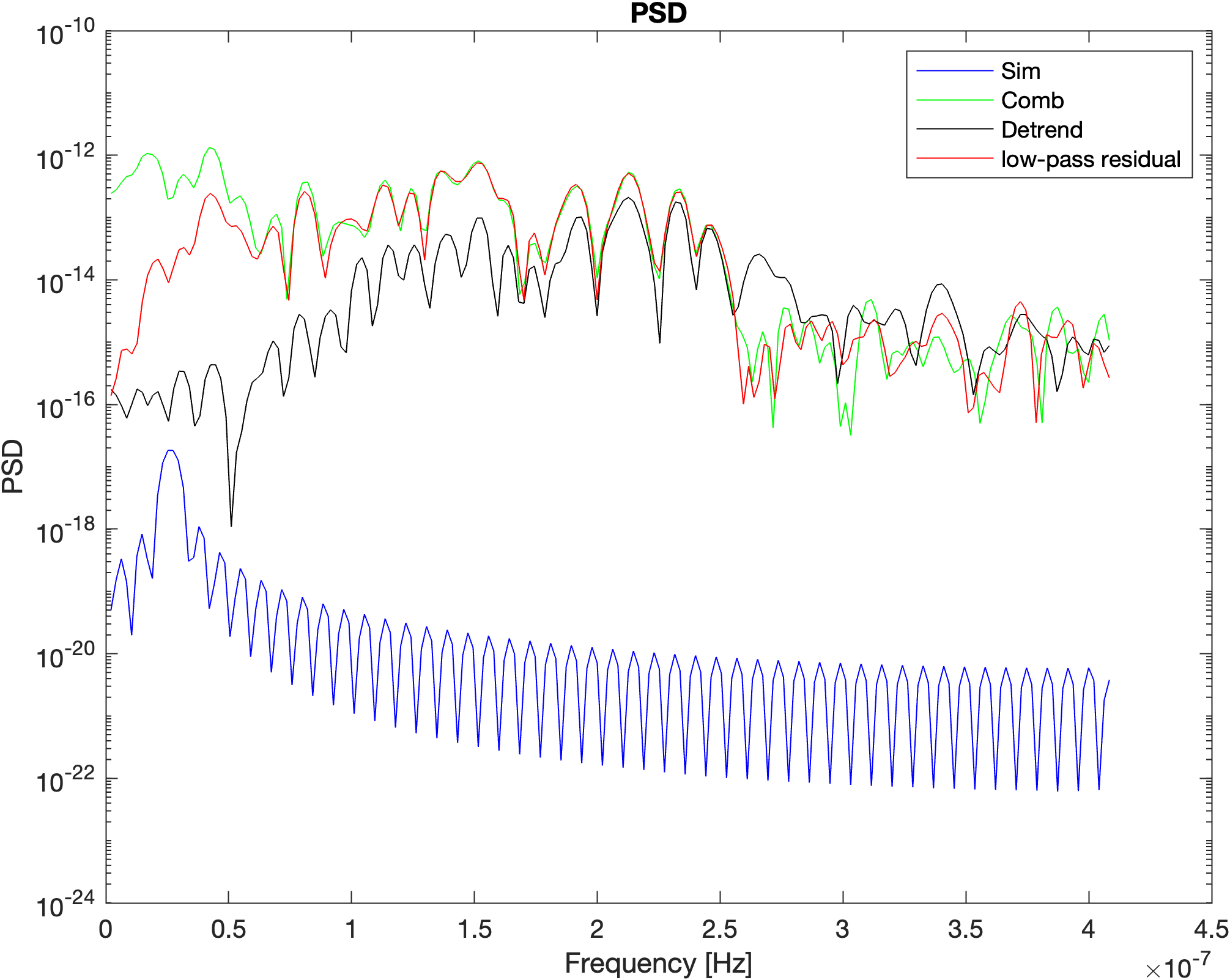}
  }

  \vspace{0.5cm} 

  \subfigure[PSR J1745+1017]{
    \label{fig:app_J1745+1017_LSP}
    \includegraphics[width=0.45\textwidth]{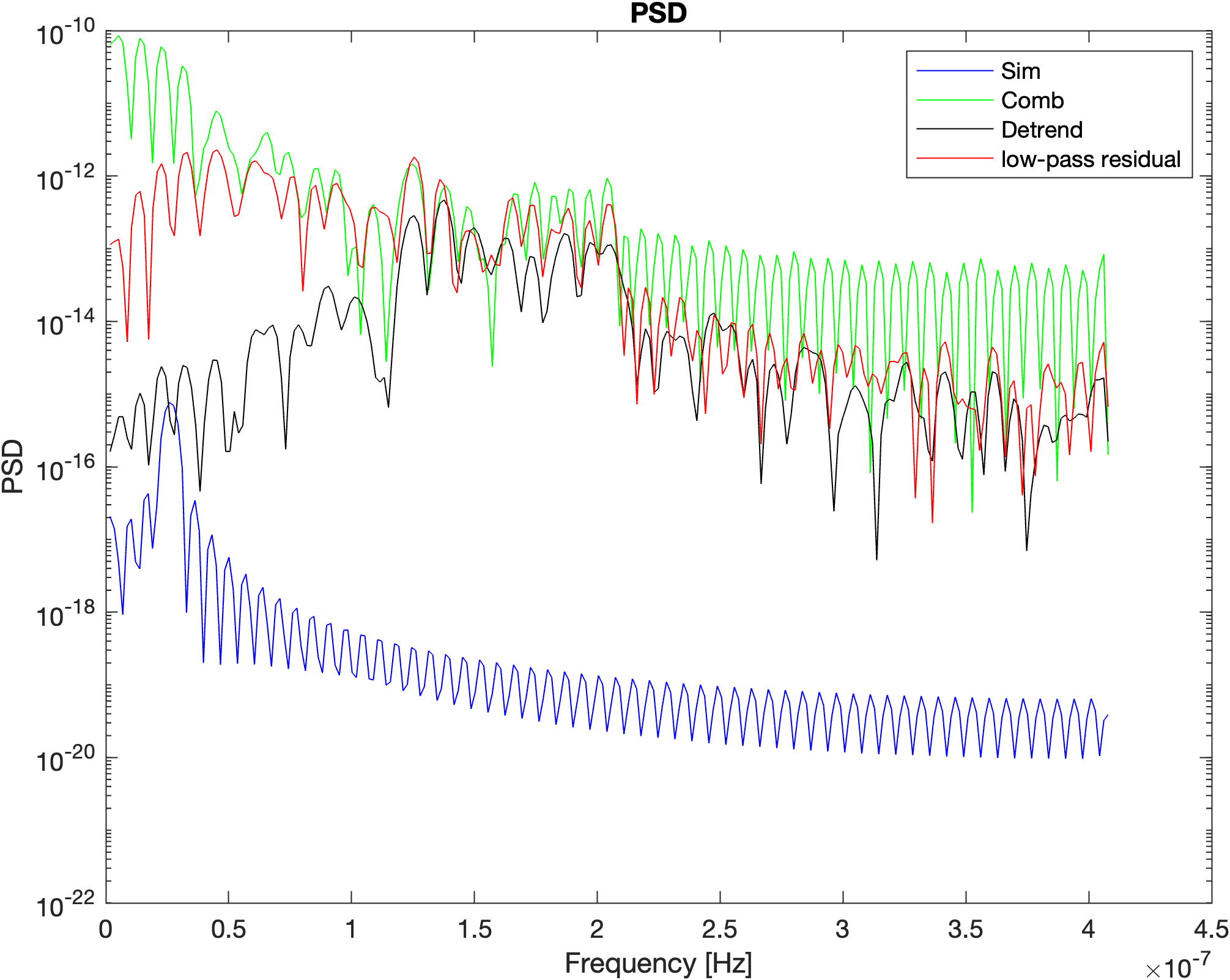}
  }
  \hfill
  \subfigure[PSR J1903+0327]{
    \label{fig:app_J1903+0327_LSP}
    \includegraphics[width=0.45\textwidth]{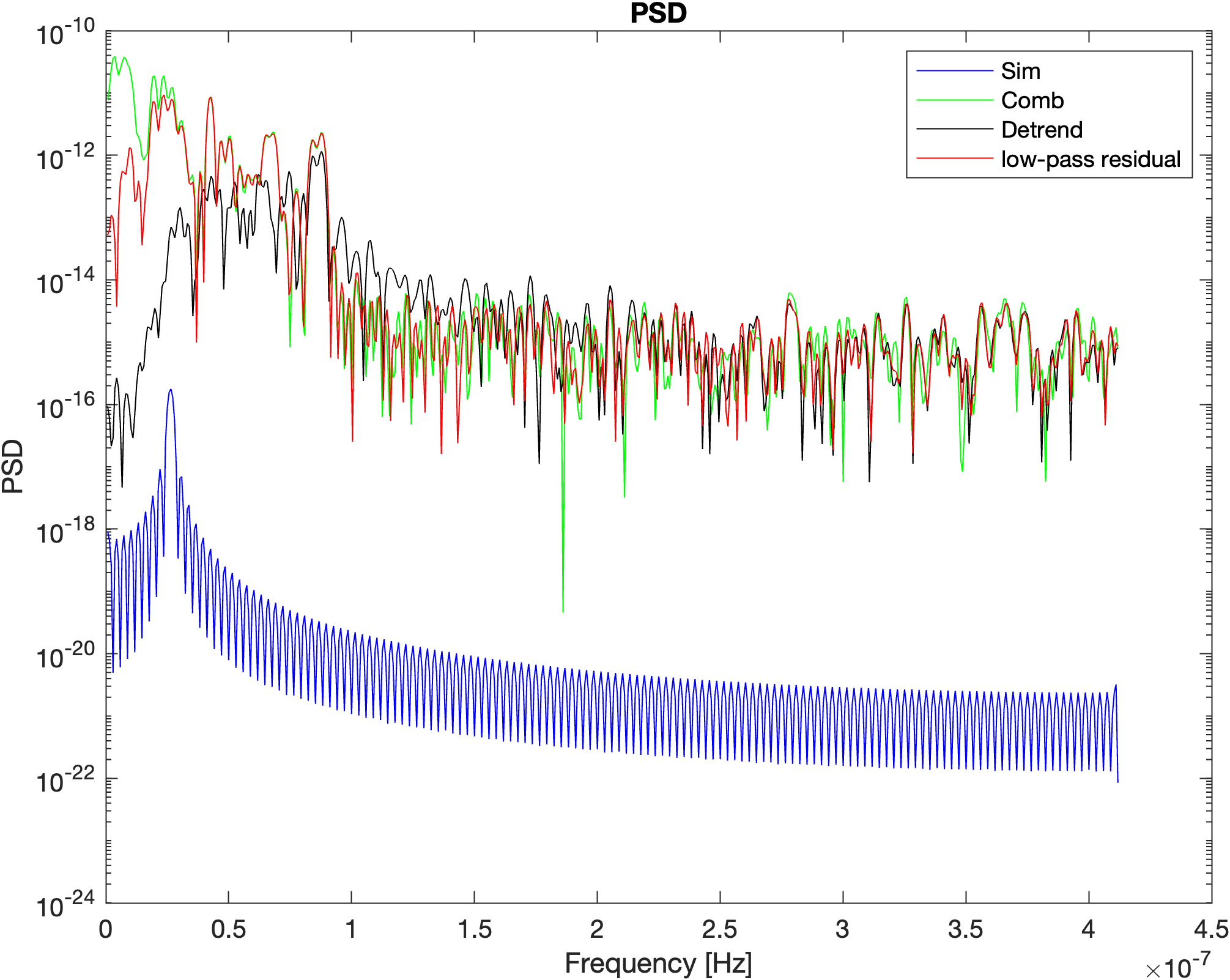}
  }

  \caption{The Lomb-Scargle periodograms for the pulsars excluded by the selection scheme. The blue line represents the injected CGW signal, the green line shows the simulated data with the injected signal, the black line indicates the \texttt{SHAPES} detrended residuals, and the red line shows the result after applying the low-pass filter. }
  \label{fig:appendix_LSPs}
\end{figure*}

In Fig.~\ref{fig:appendix_LSPs}, we present the Lomb-Scargle periodograms for the additional pulsars analyzed in this work that were excluded by the selection scheme. The poor performance of both the Bayesian and SM methods for the full PTA dataset may be accounted for by these pulsars.

\clearpage


%

\end{document}